\documentclass[11pt,a4paper]{article}
\usepackage{float}
\usepackage{multirow}
\usepackage{graphics}
\usepackage[cp1252,utf8]{inputenc}
\usepackage{hyperref}
\usepackage{graphicx} 
\usepackage{amsmath}
\usepackage{amsfonts}
\usepackage{amssymb}
\usepackage{bigints} 
\usepackage{relsize} 
\usepackage[top=1in, bottom=2cm, left=1.5cm, right=1.5cm]{geometry}
\usepackage{authblk}

\usepackage{changepage} 

\usepackage{afterpage}

\usepackage{placeins}

\usepackage{wrapfig}
\usepackage{subfigure} 

\title{\bf Investigation of circular geodesics in a rotating charged black hole in the presence of perfect fluid dark matter}
\author[a]{\bf  Anish Das \thanks{anishdas1995@bose.res.in}}
\author[b]{\bf  Ashis Saha \thanks{ashisphys18@klyuniv.ac.in}}
\author[a]{\bf Sunandan Gangopadhyay \thanks{sunandan.gangopadhyay@bose.res.in}}

\affil[a]{\textit{Department of Theoretical Sciences, S.N.~Bose National Centre for Basic Sciences,}\textit{JD Block, Sector-III, Salt Lake, Kolkata 700106, India}}
\affil[b]{\textit{Department of Physics, University of Kalyani, Kalyani 741235, India}}

\date{}

\begin{document}

\maketitle

\begin{abstract}
\noindent In this work we have obtained a charged black hole solution in the presence of perfect fluid dark matter (PFDM) and discuss its energy conditions. The metric corresponding to the rotating avatar of this black hole solution is obtained by incorporating the Newman-Janis algorithm. We then compute two types of circular geodesics, namely, the null geodesics and time-like geodesics for this rotating spacetime geometry. For the  case of time-like geodesics, we consider both neutral as well as charged massive particles. The effective potentials of the corresponding circular geodesics has also been studied. We then present our results by graphically representing the collective effects of the black hole parameters, namely, the charge of the black hole ($Q$), spin parameter ($a$) and the PFDM parameter ($\alpha$) on the energy ($E$), angular momentum ($L$) and effective potential ($V_{eff}$) of the concerned particle. Finally, we discuss the Penrose process in order to study the negative energy particles having possible existence within the ergosphere, and which in turn leads to the energy gain of the emitted particle.
\end{abstract}


\section{Introduction}
The mathematical and observational study of black holes, one of the most fascinating objects in the universe has been one of the most important areas of research since time immemorial. Although the general theory of relativity is still not able to grasp the full story of these objects, it turns out that to a great extent the  theory of general relativity is able to match the observational aspects of black holes, and is also able to predict some remarkable properties of these objects. The supermassive black holes are believed to be the central core of all the galaxies in the universe and are responsible for holding the entire galaxy together. Recently, the shadow images of  $M87$ \cite{1} provided the first direct evidence for the existence of black holes and this motivates us to study some interesting aspects of these remarkable objects called  black holes. It is difficult to study black holes directly and measure any of its properties like Hawking temperature.  So the study of black hole spacetime structure and thereby the geodesic motion of various types of particles in its vicinity, helps to gain a deeper understanding about the features of black holes. A vast amount of study has been done on the geodesic motion around black holes and it is very hard to mention all of them, yet some of them has to be looked upon due to their novelty. In \cite{2,3}, the study of geodesics around Schwarzchild and Kerr black holes can be found in details. Besides that, geodesic motion of massive particles have been studied around conformal Schwarzchild black hole in \cite{4}. The study of geodesics in regular Hayward black hole \cite{5}, Kerr-Sen black hole \cite{6,7}, Kerr-Newman-Taub-NUT black hole \cite{8}, generic black hole coupled to non-linear electrodynamics \cite{9} and Schwarzchild black hole in quintessence \cite{10} are also worth mentioning. Some interesting studies on black holes in $f(T)$ gravity and teleparallel gravity has been carried out in \cite{7a,7b}. Studies on phantom RN-$AdS$ black hole can be found in \cite{7c}. Since the observer in reality is far away from the black hole, so theoretically one places the observer at infinity from which all planes are identical and hence for convenience, the geodesics in the equatorial plane of the black holes are able to capture the whole story. The geodesics in equatorial plane for dyonic Kerr-Newman black hole has been studied in \cite{11} and that for quintessential rotating black hole with and without cosmological constant ($\Lambda$) has been studied in \cite{12} and \cite{13} respectively. Also the study of geodesics for distorted static black hole can be found in \cite{14} along with particle motion in Kerr spacetime in \cite{15} and Kerr spacetime pierced by cosmic string in \cite{16}. Also recent study of timelike geodesics in black hole spacetime can be found in \cite{1a,1b}. Besides massive test particles, the study of null geodesics demands more attention since they consume an extensive area of research now-a-days after the discovery of black hole shadow image recently by \textit{The Event Horizon Telescope} collaboration \cite{1}. The motivation to study the null geodesics is to get a complete understanding about the photon sphere which in turn leads to the formation of the black hole shadow. There are many studies on the null geodesics and thereby on the shadow of black holes. Some of them can be found  in \cite{7},\cite{17}-\cite{24}. Apart from the study of photons and massive particles, researchers have shown considerable interest in the study of charged particles around black hole spacetime due to the presence of the Maxwell fields leading to additional interacting terms. The study of charged particles around Kerr-Newman black hole has been carried out in \cite{25}-\cite{28}. Also the motion of the charged particles in case of black holes immersed in external electromagnetic fields has been studied in \cite{29}-\cite{33}.

\noindent Another fascinating thing in this context is the presence of dark matter and dark energy in our universe. Dark matter is supposed to fill about $27\%$ of the universe along with dark energy ($68\%$) and ordinary matter($5\%$). Due to the indirect observations indicating the presence of dark matter, it is believed to be present everywhere within the observable universe and the amount increases as we move away from the galactic center. This is inferred \textit{from the observation of galaxy rotation curves}. There are different models of dark matter which are very useful to explain large and small scale structure of the universe \cite{34}-\cite{36}. 
Besides that, in most theories dark matter is assumed to be present around black holes as \textit{halos} and black hole shadows has been calculated in those cases  \cite{37,38}. Also quintessential dark matter solutions were studied by Kiselev \cite{39}. In recent times dark matter has also been studied using a perfect fluid model \cite{40}. This interesting model helps to investigate the impact of dark matter on various observables related to black holes. There has been study of shadows in perfect fluid dark matter (PFDM) in case of rotating black holes with and without cosmological constant \cite{41,42} along with non-rotating charged black holes \cite{43}. Dark matter is composed of non-baryonic matter and being present around black holes can contribute to the effective mass of the black hole system. The geodesics around black holes in PFDM should in principle have some intriguing features due to the presence of extra matter and hence can lead to interesting consequences. Besides that, as theorised by Penrose \cite{44}, it is well known that black holes are sources of extreme energy. The process has maximum efficiency at the event horizon of the black hole and requires particles of negative energy and angular momentum to be absorbed by the black hole which requires the presence of a region known as the \textit{ergosphere}, the place where the process takes place. There are different studies on the Penrose process and thereby the extraction of energy from black holes \cite{45}-\cite{48}, where energy extraction for spinning particles have been studied, and also it has been found that the efficiency of the Penrose process increases in higher dimensions. In this paper we study the Penrose process for Kerr-Newman black holes in asymptotically flat spacetime in the presence of PFDM. We briefly describe the work carried out in this paper in the following paragraph.

\noindent In this work, we obtain a static, charged black hole solution surrounded by PFDM and look at its energy conditions. We obtain the rotating version of the black hole solution by incorporating the Newman-Janis algorithm. The variation of the event horizon with respect to the PFDM parameter $\alpha$ has been observed. We then study the geodesics corresponding to both massless and massive particles in the vicinity of the rotating charged black hole with PFDM. We then study the circular geodesics of photons. Furthermore, we look into how the dark matter parameter $\alpha$ as well as the spin ($a$) and charge ($Q$) of the black hole affect the photon radius. In case of massive particles, we compute the energy ($E$) and the angular momentum ($L$) of the particle in addition to the effective potential ($V_{eff}$). The effects of the PFDM parameter $\alpha$, spin ($a$) and charge ($Q$) on these quantities, has been displayed in graphs and Tables. We then study the Penrose process. Our aim here is to investigate  how the PFDM parameter $\alpha$ affects the size of the ergosphere, the negative energy of the particle and thereby the efficiency of the Penrose process.

\noindent The paper is organised as follows. In section \ref{sec2}, we derive the charged black hole spacetime in the presence of PFDM and discuss its energy conditions. In section \ref{sec3}, we obtain the rotating avatar of the black hole solution by using the Newman-Janis algorithm. In section \ref{sec4}, we discuss the circular geodesics of various particles (photons, uncharged and charged massive particles) and then we show the nature of  the potential of the black hole spacetime in section \ref{sec5}. Then we discuss the Penrose process in detail in section \ref{sec6} and conclude by summarising our results in section \ref{sec7}. This paper is supplemented by two Appendices. We have assumed $\hbar=c=G=1$ throughout this work.


\section{Charged black hole in perfect fluid dark matter}
We consider a ($3+1$)-dimensional gravity theory minimally coupled with a U(1) gauge field, in the presence of perfect fluid dark matter (PFDM). The action can be written in the following form \cite{36,39,Kiselev1,PEDM1,43}
\label{sec2}\begin{equation}
S=\int d^4 x\sqrt{-g}\Big(\frac{R}{16\pi G} + \frac{1}{4}F^{\mu \nu}F_{\mu \nu} +\mathcal{L}_{DM}\Big)
\end{equation}
where $R$ is the Ricci scalar and $G$ is the Newton's gravitational constant. $F_{\mu \nu}$ is the Maxwell field strength which is related to the electromagnetic potential $A_{\mu}$ as $F_{\mu \nu}=\partial_{\mu}A_{\nu}-\partial_{\nu}A_{\mu}$ and $\mathcal{L_{DM}}$ gives the Lagrangian density for the perfect fluid dark matter. Extremizing the action we obtain the Einstein field equations as
\begin{equation}\label{1201}
R_{\mu \nu}-\frac{1}{2}g_{\mu \nu}R=8\pi G( T_{\mu\nu}^{M}-T_{\mu\nu}^{DM})\equiv - 8\pi G T_{\mu\nu}~.
\end{equation}
In the above equation, $ T_{\mu\nu}^{M}$ represents the energy-momentum tensor corresponding to the ordinary matter (in this case the Maxwell field). This can be denoted as \cite{55}
\begin{eqnarray}
(T^{\mu}_{~\nu})^{M}=diag\left(-\frac{Q^2}{8\pi G r^4}, -\frac{Q^2}{8\pi G r^4}, \frac{Q^2}{8\pi G r^4}, \frac{Q^2}{8\pi G r^4}\right)
\end{eqnarray}
where $Q$ is the electric charge. On the other hand, $ T_{\mu\nu}^{DM}$ corresponds to the energy-momentum tensor of the perfect fluid dark matter \cite{49}. It is specified as 
\begin{eqnarray}\label{A1}
(T^{\mu}_{~\nu})^{DM}=diag(-\rho, P_r, P_{\theta}, P_{\phi})~;~P_r = -\rho~;~P_{\theta}=P_{\phi}=P
\end{eqnarray}
where $\rho$ and $P$ correspond to density and pressure of the perfect fluid dark matter. Following the approach given in \cite{PEDM1,49}, we further consider  
\begin{eqnarray}\label{A2}
(T^{\theta}_{~\theta})^{DM} = (T^{t}_{~t})^{DM}(1-\epsilon)~~;~~(T^{\phi}_{~\phi})^{DM} = (T^{r}_{~r})^{DM}(1-\epsilon)
\end{eqnarray}
with $\epsilon$ being a constant. Hence, the total energy-momentum tensor reads
\begin{eqnarray}
	T_{~\nu}^{\mu} = diag\left(-\rho_s, P_r^{total}, P_{\theta}^{total}, P_{\phi}^{total}\right)
\end{eqnarray}
where $\rho_s= \rho - \frac{Q^2}{8\pi G r^4}$, $P_r^{total}= - P_{r}- \frac{Q^2}{8\pi G r^4}$, $P_{\theta}^{total}= -P+ \frac{Q^2}{8\pi G r^4}$ and $P_{\phi}^{total}=-P+\frac{Q^2}{8\pi G r^4}$.\\

\noindent By substituting  eq.(\ref{A1}) in eq.(\ref{A2}), we obtain the equation of state for the PFDM to be \cite{49}
\begin{eqnarray}\label{A3}
\frac{P}{\rho}=(\epsilon - 1)~.
\end{eqnarray}
In order to obtain a static, spherically symmetric solution, we assume an ansatz metric of the form
\begin{equation}
ds^2 =-e^{\nu}dt^2 + e^{\lambda}dr^2 + r^2 (d\theta ^2 + \sin ^2 \theta d\phi^2)
\end{equation}
with $\nu$, $\lambda$ being functions of $r$ only. Now the Einstein field equations read
\begin{eqnarray}
e^{-\lambda}\Big(\frac{1}{r^2}-\frac{\lambda^{'}}{r}\Big)-\frac{1}{r^2}=8\pi G \Big(\rho-\frac{Q^2}{8\pi G r^4}\Big)\equiv 8\pi G\rho_s\label{EoMs1}\\
e^{-\lambda}\Big(\frac{1}{r^2}+\frac{\nu^{'}}{r}\Big)-\frac{1}{r^2}=8\pi G \Big(-P_r-\frac{Q^2}{8\pi G r^4}\Big)\equiv -8\pi G P_r^{total}\label{EoMs2}\\
\frac{e^{-\lambda}}{2}\Big(\nu{''}+\frac{\nu{'}^2}{2}+\frac{\nu{'}-\lambda^{'}}{r}-\frac{\nu^{'}\lambda^{'}}{2}\Big)=8\pi G \Big(-P+\frac{Q^2}{8\pi G r^4}\Big)\equiv- 8\pi G\ P_{\theta}^{total}\label{EoMs3}\\
\frac{e^{-\lambda}}{2}\Big(\nu{''}+\frac{\nu{'}^2}{2}+\frac{\nu{'}-\lambda^{'}}{r}-\frac{\nu^{'}\lambda^{'}}{2}\Big)=8\pi G \Big(-P+\frac{Q^2}{8\pi G r^4}\Big)\equiv- 8\pi G\ P_{\phi}^{total}\label{EoMs4}
\end{eqnarray}
where the prime denotes derivative with respect to the radial coordinate. Eq.(\ref{EoMs1}) and eq.(\ref{EoMs3}) can be rearranged to the form
\begin{eqnarray}\label{1206}
e^{-\lambda}\Big(\frac{1}{r^2}-\frac{\lambda^{'}}{r}\Big)-\frac{1}{r^2}+\frac{Q^2}{ r^4}=8\pi G \rho \nonumber\\
\frac{e^{-\lambda}}{2}\Big(\nu{''}+\frac{\nu{'}^2}{2}+\frac{\nu{'}-\lambda^{'}}{r}-\frac{\nu'\lambda^{'}}{2}\Big)-\frac{Q^2}{ r^4}=-8\pi G P~. 
\end{eqnarray}
By using the equation of state for the PFDM (given in eq.(\ref{A3})) in the above equations and taking their ratio, we get
\begin{equation}\label{1203}
\frac{e^{-\lambda}}{2}\Big(\nu{''}+\frac{\nu{'}^2}{2}+\frac{\nu{'}-\lambda^{'}}{r}-\frac{\nu'\lambda^{'}}{2}\Big)-\frac{Q^2}{ r^4}=(1-\epsilon)\Bigg[e^{-\lambda}\Big(\frac{1}{r^2}-\frac{\lambda^{'}}{r}\Big)-\frac{1}{r^2}+\frac{Q^2}{ r^4}\Bigg]~.
\end{equation}
Now subtracting eq.(\ref{EoMs2}) from eq.(\ref{EoMs1}), we get
\begin{eqnarray}
	\nu^{\prime} + \lambda^{\prime} &=&0\nonumber\\
	\nu + \lambda &=& \mathcal{C}~.
\end{eqnarray}
Rescaling the time coordinate, the constant of integration $\mathcal{C}$ can be set to zero. This implies
\begin{eqnarray}
	\nu = - \lambda~.
\end{eqnarray}
Setting $\nu=\ln(1-U)$, where $U\equiv U(r)$ simplifies eq.(\ref{1203}) to the following form
\begin{equation}\label{1204}
\frac{U^{''}}{2}+\epsilon\frac{U}{r} + (\epsilon -1)\frac{U}{r^2} + (2-\epsilon)\frac{Q^2}{r^4}=0~.
\end{equation}
Eq.\eqref{1204} can be solved for different values of $\epsilon$ \cite{PEDM1}. However we are particularly interested in the solution for $\epsilon=\frac{3}{2}$ \cite{Kiselev1,PEDM1}. For $\epsilon=\frac{3}{2}$, eq.(\ref{1204}) reduces to the following form
\begin{equation}
r^2 U^{''}+3rU^{'} + U + \frac{Q^2}{r^2}=0~.
\end{equation}
The solution of the above equation is obtained to be
\begin{equation}\label{u}
U(r)=\frac{r_s}{r}-\frac{Q^2}{r^2}-\frac{\alpha}{r}\ln\Big(\frac{r}{|\alpha|}\Big)
\end{equation}
where $r_s$ and $\alpha$ are integration constants. In order to evaluate $r_s$, we set $Q=0$ and $\alpha=0$. In this limit, by utilizing the weak field approximation, $r_s$ is obtained to be $2GM$. The lapse function therefore becomes
\begin{eqnarray}\label{solution}
f(r)= e^{\nu}=e^{-\lambda}=e^{\ln(1-U)}=1-U=1-\frac{2GM}{r}+\frac{Q^2}{r^2}+\frac{\alpha}{r}\ln\Big(\frac{r}{|\alpha|}\Big)
\end{eqnarray}
corresponding to the following metric of a static, spherically symmetric, charged black hole in PFDM
\begin{equation}\label{sol1}
ds^2 =-f(r)dt^2 +\frac{1}{f(r)}dr^2 + r^2 d\theta^2 + r^2\sin^2 \theta  d\phi^2~.
\end{equation}
It is reassuring to observe that the above geometry reduces to the well-known black hole solutions in the literature. In the limit $\alpha\rightarrow 0$, eq.(\ref{sol1}) coincides with the Reissner-Nordstr\"om black hole solution and in the limit $\alpha\rightarrow 0, Q \rightarrow 0$ it reduces to the Schwarzschild solution. We have shown these limiting cases in Appendix A. Further, in the limit $Q\rightarrow0$ the above geometry matches with the solution given in \cite{42,49}.

\subsection{Energy conditions}
We now comment on the energy conditions of the above black hole solution. Firstly, we shall derive the density ($\rho$) of the dark matter. This can be done by substituting $e^{-\lambda}$ given in eq.(\ref{solution}), in eq.(\ref{EoMs1}). This yields
\begin{eqnarray}
\rho = \frac{\alpha}{8\pi G r^3}~.
\end{eqnarray}
Similarly, by using the rest of the equations, we can obtain the pressure components of the dark matter energy-momentum stress tensor $(T^{\mu}_{~\nu})^{DM}$. This reads
\begin{eqnarray}
P_r = -\rho = - \frac{\alpha}{8\pi G r^3},~ P_{\theta}= P_{\phi} = \frac{1}{2} \rho~.
\end{eqnarray}
The weak energy condition (WEC) can be written down as \cite{Carroll,Wald}
\begin{eqnarray}\label{WEC}
	T_{\mu\nu}\xi^{\mu}\xi^{\nu}\geq 0
\end{eqnarray}
where $\xi^{\mu}$ and $\xi^{\nu}$ are time-like vectors.
 This energy condition signifies that the total energy density of all matter fields measured by an observer traversing a time-like curve is always positive. The WEC (\ref{WEC}) implies $\rho_s\geq0$ and $\rho_s+P_i^{total}\geq 0~(i=r,\theta,\phi)$ (where $P_i^{total}$ is the pressure in any direction). Hence, in order to satisfy the WEC, the following condition must hold
 \begin{eqnarray}
 \rho_s &=& \rho - \frac{Q^2}{8\pi G r^4} \geq 0\nonumber\\
 &=& \frac{\alpha}{8\pi G r^3} - \frac{Q^2}{8\pi G r^4}\geq 0.
 \end{eqnarray}
The above condition reduces to $\frac{\alpha}{8\pi G r^3}\geq 0$, that is, 
$\alpha\geq 0$ when the (ordinary) matter field is absent (that is $Q=0$).
 

\section{Rotating charged black hole in perfect fluid dark matter: Newman-Janis algorithm}\label{sec3}
From the point of view of a more realistic set up, inclusion of spin parameter in the black hole metric is necessary. In order to incorporate spin ($a$), we shall use the Newman-Janis algorithm \cite{50}, \cite{57} which is by far the easiest and most effective technique to include the spin parameter to any metric without cosmological constant. Before that we would like to consider a general metric and follow the approach accordingly as in \cite{51}. Let us assume a metric of the form
\begin{equation}\label{33}
ds^2 =-f(r)dt^2 +\frac{1}{g(r)}dr^2 + h(r)\Big(d\theta^2 + \sin^2 \theta  d\phi^2 \Big)~.
\end{equation}
Then the metric can be rewritten in the Eddington-Finkelstein coordinates ($u$) by the transformation
\begin{equation}
dt=du + \frac{dr}{\sqrt{f(r)g(r)}}
\end{equation}
which modifies the metric as
\begin{equation}
ds^2 =-f(r)du^2 -2\sqrt{\frac{f(r)}{g(r)}}du dr + h(r)\Big(d\theta^2 + \sin^2 \theta  d\phi^2 \Big)~.
\end{equation}
We now introduce the null tetrads $Z^\mu = \Big(l^\mu,n^\mu,m^\mu,\overline{m}^\mu\Big)$ in terms of which the metric tensor can be written as
\begin{equation}
g^{\mu \nu}=-l^\mu n^\nu -l^\nu n^\mu + m^\mu \overline{m}^\nu + m^\nu \overline{m}^\mu
\end{equation}
with the relation between the tetrad components as ~$l^\mu n_\mu = -m^\mu \overline{m}_\mu =1$ and all others are zero. Now in order to get the inverse metric, we need to represent the tetrad components in terms of ($u,r,\theta,\phi$). This reads
\begin{equation}\nonumber
l^\mu =\delta ^\mu _r
\end{equation}
\begin{equation}\nonumber
n^\mu =\sqrt{\frac{g(r)}{f(r)}}\delta ^\mu _u -\frac{g(r)}{2}\delta^\mu _r
\end{equation}
\begin{equation}
m^\mu =\frac{1}{\sqrt{2h(r)}}\Big(\delta^\mu _\theta + \frac{i}{\sin \theta}\delta^\mu _\phi\Big)~.
\end{equation}

\noindent To incorporate spin to the metric, we make the transformation of the form $u\to u^{'}=u-ia\cos \theta$ and $r \to r^{'}=r + ia\cos \theta$ which transforms the null tetrads, and the transformation relations look like
\begin{equation}
Z^{'\mu}_\beta = \frac{\partial x^{' \mu}}{\partial x^{\nu}}Z^{\nu}_\beta~.
\end{equation}
Here $\mu$ denotes the components of the tetrads along the directions $u,r,\theta,\phi$ and $\beta$ denotes the tetrads ($l,n,m,\bar{m}$). So the transformation results in a new set of tetrads which read 
\begin{equation}\nonumber
l^{'\mu} =\delta ^\mu _r;~~n^{'\mu} =\sqrt{\frac{g(r)}{f(r)}}\delta ^\mu _u -\frac{f(r)}{2}\delta^\mu _r;~~m^{'\mu} =\frac{1}{\sqrt{2h(r)}}\Big(ia\sin \theta \Big(\delta^\mu _u - \delta^\mu _r\Big)+\delta^\mu _\theta + \frac{i}{\sin \theta}\delta^\mu _\phi\Big)~.
\end{equation}
Since $r$ and $u$ got transformed, hence the components of the metric tensor will also change. So the modified functions are $f(r) \to F(r,\theta)$, $g(r) \to G(r,\theta)$ and $h(r) \to H(r,\theta)$. The non-zero components of the inverse metric tensor are as follows
\begin{eqnarray}\nonumber
g^{uu} =\frac{a^2 \sin^2 \theta}{H}~;~g^{rr}=G+\frac{a^2 \sin^2 \theta}{H}~;~	g^{ur} =	g^{ru}=-\sqrt{\frac{G}{F}}-\frac{a^2 \sin^2 \theta}{H}
\end{eqnarray}
\begin{eqnarray}\nonumber
g^{\theta \theta}=\frac{1}{H}~;~g^{\phi \phi}=\frac{1}{H \sin^2 \theta}~;~	g^{u \phi}=g^{\phi u}=\frac{a}{H}~;~g^{r \phi}=g^{ \phi r}=-\frac{a}{H}~.
\end{eqnarray}
We now obtain the non-zero components of the metric
\begin{eqnarray}\nonumber
g_{uu} =-F~;~g_{ur}=g_{ru}=-\sqrt{\frac{F}{G}}~;~g_{u \phi}=g_{\phi u}=\Big(-a\sqrt{\frac{F}{G}}+aF\Big)\sin ^2 \theta
\end{eqnarray}
\begin{eqnarray}\nonumber
g_{\theta \theta}=H~;~g_{\phi \phi}= \sin^2 \theta\Big(H+ 2a^2 \sin^2 \theta\sqrt{\frac{F}{G}}-a^2 F \sin^2 \theta \Big)~;~g_{r \phi}=g_{\phi r}=a \sin^2 \theta\sqrt{\frac{F}{G}}~.
\end{eqnarray}
The resulting metric reads
\begin{eqnarray}
ds^2 = -F du^2 - 2\sqrt{\frac{F}{G}}dudr + 2a\sin^2 \theta \Big(F-\sqrt{\frac{F}{G}}\Big)dud\phi + 2a\sin^2 \theta \sqrt{\frac{F}{G}}drd\phi \nonumber \\ 
+ H d\theta^2 + 
\sin^2 \theta \Big[H+a^2 \sin^2 \theta \Big(2\sqrt{\frac{F}{G}}-F\Big)\Big]d\phi^2~.
\end{eqnarray}
Now we need to remove $u$ and express the metric in the Boyer-Lindquist coordinates for which the necessary transformation is $du=dt + \xi_{1} (r)dr$ and $d\phi=d\phi + \xi_{2} (r)dr$. Only the diagonal elements and $dtd\phi$ component of the metric survives, rest are zero. This leads to the values of $\xi_{1}$ and  $\xi_{2}$ as
\begin{equation}\nonumber
\xi_{1}(r)=\frac{-\Big(\sqrt{\frac{G}{F}}H +a^2 \sin^2 \theta \Big)}{\Big(GH +a^2 \sin^2 \theta\Big)}~~;~~\xi_{2}(r)=\frac{-a}{\Big(GH +a^2 \sin^2 \theta\Big)}
\end{equation}
and the metric becomes
\begin{eqnarray}
ds^2 = -F dt^2 +\Big(\frac{H}{GH +a^2 \sin^2 \theta}\Big)dr^2 + H d\theta^2+ 2a\sin^2 \theta \Big(F-\sqrt{\frac{F}{G}}\Big)dtd\phi  \nonumber \\ 
+ 
\sin^2 \theta \Big[H+a^2 \sin^2 \theta \Big(2\sqrt{\frac{F}{G}}-F\Big)\Big]d\phi^2~.
\end{eqnarray}
\noindent Now in order to keep the terms in the metric real, we use the transformation  $r^{p}\to \frac{r^{p+2}}{r^2 + a^2 \cos^2 \theta}=\frac{r^{p+2}}{\rho^2} $ ; $p\geq 0$ following \cite{52}. This approach of incorporating spin ($a$) into the terms, leads to the modified functions of the form
\begin{equation}\nonumber
f(r) \to F(r)=1 -\frac{2Mr}{\rho^2}+ \frac{Q^2}{\rho^2} + \frac{\alpha r}{\rho^2} ln\Big(\frac{r}{|\alpha|}\Big)
\end{equation}
\begin{equation}\nonumber
g(r) \to G(r)=1 -\frac{2Mr}{\rho^2}+ \frac{Q^2}{\rho^2} + \frac{\alpha r}{\rho^2} ln\Big(\frac{r}{|\alpha|}\Big)
\end{equation}
\begin{equation}\nonumber
h(r) \to H(r)=\rho^2~.
\end{equation}
Using the above relations, we obtain the final expression for the metric of rotating charged black hole in PFDM to be
\begin{align}
ds^2 =-\frac{1}{\rho^2}\Big(\Delta -a^2 \sin^2 \theta\Big)dt^2 +\frac{\rho^2}{\Delta}dr^2 + \rho^2 d\theta^2 -\frac{2a\sin^2 \theta}{\rho^2}\Big[2Mr -Q^2 -\alpha r \ln\Big(\frac{r}{|\alpha| }\Big)\Big] dtd\phi \nonumber \\
+ \sin^2 \theta \Big[r^2 + a^2 + \frac{a^2 \sin^2 \theta}{\rho^2}\Big(2Mr -Q^2 -\alpha r \ln\Big(\frac{r}{|\alpha| }\Big)\Big)\Big]d\phi^2	 
\end{align}
with ~~ $\Delta=r^2 +a^2 -2Mr +Q^2+\alpha r \ln\Big(\frac{r}{|\alpha| }\Big) $ and ~ $\rho^2 = r^2 + a^2 \cos^2 \theta$.
\section{Circular geodesics}\label{sec4}
As we have mentioned previously, we are interested in an observer far away (theoretically at infinity) from the black hole. This motivates us to confine ourselves only for the case of geodesics in the equatorial plane ($\theta=\frac{\pi}{2}$). The consideration of equatorial plane simplifies the following functions as  
\begin{equation}\nonumber
\Delta=r^2 +a^2 -2Mr +Q^2+\alpha r \ln\Big(\frac{r}{|\alpha|}\Big)~~;~~ \rho^2 =r^2~.
\end{equation}
The logarithmic term with parameter $\alpha$ corresponds to the contribution of the PFDM to the charged black hole metric. This term can be positive or negative depending on the sign of $\alpha$. In \cite{49}, the authors have discussed about relevance of both positive and negative values of $\alpha$ but here we are interested only in positive values ($\alpha>0$). The allowed maximum value for $\alpha$ is $\alpha_{max}=2M$ \cite{41,49}. For the sake of simplicity, we will consider $M=1$ in our subsequent analysis. The analysis with negative values of $\alpha$ can also be done by following the subsequent analysis. The PFDM term influences the structure of the spacetime and hence the trajectories of the geodesics.
\begin{center}
	\begin{table}[h]
		\centering
		\begin{tabular}{|c|c|c|c|c|c|c|c|c|c|c|c|} 	 	
			\multicolumn{11}{c}{\textbf{}}\\
			\hline
			$\alpha$ & 0.1 & 0.2 & 0.3 & 0.4 & 0.47 & 0.5 & 0.6  & 0.7 & 0.8 & 0.9 & 1.0  \\
			\hline
			$r_{h+}$ &   1.5 & 1.37 & 1.30 & 1.27 & 1.265 & 1.267 & 1.28 & 1.30 & 1.34 & 1.37 & 1.415 \\ 
			\hline
			$r_{h-}$ & 	 0.2 & 0.185 & 0.17 & 0.1525 & 0.14 & 0.135 & 0.12 & 0.105 & 0.095 & 0.085 & 0.075   \\
			\hline	
		\end{tabular}
		\caption{\small Values of inner ($r_{h-}$) and outer ($r_{h+}$) horizon of the black hole for various values of $\alpha$ at $a=0.5,~Q=0.3$.}
		\label{322}
	\end{table}
\end{center}
In Table \ref{322}, we show the variations in the values of the inner horizon ($r_{h-}$) and outer horizon ($r_{h+}$) of the black hole for various values of $\alpha$ at constant spin parameter $a =0.5$ and charge $Q =0.3$ of the black hole. We find that the outer horizon ($r_{h+}$) decreases with the increase in the value of $\alpha$. This behavior can be observed upto certain critical value $\alpha_c$ (which in this case $\alpha_c\approx0.47)$, however, after this critical value ($\alpha_c$), the outer horizon ($r_{h+}$) starts to increase slowly. On the other hand, with the increase in the value of $\alpha$, the value of the inner horizon ($r_{h-}$) decreases. This critical value $\alpha_c$ can be interpreted as the point of reflection. The point of reflection can be found by plotting $\Delta(r)$ with r at fixed values of spin ($a$) and charge ($Q$) of the black hole. With increase in $\alpha$, the outer event horizon ($r_{h+}$) reduces and at some value of $\alpha$ starts to increase. That value of $\alpha$ gives the point of reflection ($\alpha_{c}$).

\begin{table}[h]\label{table}
	\centering
	\begin{tabular}{|c|c|c|} 	 	
		\multicolumn{3}{c}{}\\
		\hline
		$a$ & $Q$  & $\alpha_c$ \\
		\hline
		0.1 & 0.0 & 0.602\\ 
		\hline
		0.1   & 0.8 & 0.454    \\
		\hline
		0.2 & 0.0 & 0.596\\ 
		\hline
		0.2 & 0.8 & 0.432\\ 
		\hline
		0.3 & 0.0 & 0.574\\ 
		\hline
		0.3 & 0.8 & 0.402\\ 
		\hline
		0.4 & 0.0 & 0.552\\ 
		\hline
		0.4 & 0.7 & 0.454\\ 
		\hline
	\end{tabular}
	\hfil
	\begin{tabular}{|c|c|c|} 	 	
		\multicolumn{3}{c}{}\\
		\hline
		$a$ & $Q$  & $\alpha_c$ \\
		\hline
		0.5 & 0.0 & 0.536\\ 
		\hline
		0.5 & 0.7 & 0.418    \\
		\hline
		0.6 & 0.0 & 0.510\\ 
		\hline
		0.6 & 0.6 & 0.418\\ 
		\hline
		0.7 & 0.0 & 0.484\\ 
		\hline
		0.7 & 0.5 & 0.418\\ 
		\hline
		0.8 & 0.0 & 0.465\\ 
		\hline
		0.8 & 0.3 & 0.400\\ 
		\hline
	\end{tabular}
	\caption{\small Critical values of PFDM parameter ($\alpha_c$) for valid combinations of spin ($a$) and charge ($Q$) of the black hole. }
	\label{30}
\end{table} 

\noindent This effect has also been observed for the shadow of the rotating black hole with PFDM \cite{42}. Due to such effect observed on event horizon of black hole, we are interested in analysing different properties of the black hole spacetime in terms of the nature of particles in two ranges of $\alpha$, namely, the lower range $ \alpha < \alpha_c$ and the higher range $\alpha > \alpha_c$. This apparent increase in the size of the black hole may be assigned to the fact that after a critical value ($\alpha_c$) of the dark matter, it contributes to the effective mass of the black hole system. It can be explained by the fact that dark matter acts as a point mass distribution. So as mentioned in \cite{42}, we consider that the total system consists of two parts, one is the original BH with mass $M$ and the other part summarizing the dark matter with mass $M^{'}$. When the PFDM parameter $\alpha$ is less than the critical value $\alpha_c$, then the dark matter hinders the original black hole system, hence the effective horizon is less than $2M$. But as $\alpha$ gradually increases and becomes $\alpha > \alpha_c$, the total system is dominated by the dark matter component. Thus, the event horizon effectively increases. Hence we observe such effects in the system concerned. In Table \ref{30}, we show the values of $\alpha_c$ for various valid combinations of spin and charge which has been used in the subsequent analysis. In Table \ref{30}, we fix the spin $a$ of the black hole at a particular value, then we obtain the values of $\alpha_c$ for various values of charge $Q$. We observe that the obtained values of $\alpha_c$ lies within the range $\alpha_c \in [0.4-0.602]$ depending upon the values of spin ($a$) and charge ($Q$). On the basis of these values, we define the lower range of values of $\alpha$ which are less than these values and higher range of values of $\alpha$ which are greater than these values.

\noindent In order to continue the analysis for circular geodesics, we consider a particle with Lagrangian $\mathcal{L}=\frac{1}{2}g_{\mu \nu}\dot{x}^{\mu}\dot{x}^{\nu}$ where $\dot{x}^{\mu}=u^{\mu}=\frac{dx^{\mu}}{d\lambda}$ is the four-velocity obtained by undertaking the derivative of spacetime position ($x^{\mu}$) with respect to the affine parameter $\lambda$. The affine parameter corresponds to the proper time ($\tau$) of the massive particles in case of timelike geodesics. The Lagrangian is expressed in terms of metric and we observe that the metric coefficients are independent of $t$ and $\phi$, hence the metric is invariant along those directions (directions of symmetry) which results into conserved quantities $E$ and $L$. These two quantities physically represent the specific energy (energy per unit mass) and angular momentum (angular momentum per unit mass) of the particle respectively with respect to a stationary observer at relatively infinite distance. In terms of these quantities, the geodesic equations of $t$ and $\phi$ takes the following form
\begin{eqnarray}
\dot{t}&=&\frac{1}{r^2}\Big[\frac{r^2 +a^2}{\Delta}\Big(E(r^2 +a^2)-aL\Big)+a(L-aE)\Big]\\
\dot{\phi}&=&\frac{1}{r^2}\Big[\frac{a}{\Delta}\Big(E(r^2 +a^2)-aL\Big)+(L-aE)\Big]~.
\end{eqnarray} 
The Hamiltonian $H=p_{\mu}\dot{x}^{\mu}-\mathcal{L}$ therefore reads \begin{equation}
2H=p_{0}\dot{x}^{0}+p_{1}\dot{x}^{1}+p_{3}\dot{x}^{3}=-E\dot{t} +L\dot{\phi}+\frac{r^2}{\Delta}\dot{r}^2=constant=\epsilon~.
\end{equation}
with $\epsilon=-1,0,1$ for timelike, null and spacelike geodesics respectively. We are mainly interested in the first two types of geodesics which are physically relevant. Substituting the values of $\dot{t}$ and $\dot{\phi}$, we obtain the geodesic equation for $r$. This reads 
\begin{equation}\label{100}
\dot{r}^2 =\frac{1}{r^4}\Big[\Big(E(r^2 +a^2)-aL\Big)^2-\Delta(L-aE)^2 \Big] +\frac{\Delta}{r^2}~\epsilon~.
\end{equation}
The radial equation is very useful for the analysis of the circular geodesics and also for the computation of the effective potential.
\subsection{Null geodesics} 
In case of null geodesics $\epsilon=0$, hence the radial equation becomes
\begin{equation}\label{363}
\dot{r}^2 =\frac{1}{r^4}\Big[\Big(E(r^2 +a^2)-aL\Big)^2-\Delta(L-aE)^2 \Big] =F(r)~.
\end{equation}
For the sake of convenience we define $\frac{L}{E}=D$ as the impact parameter which reduces two constants into one. In terms of the impact parameter, the above equation becomes
\begin{equation}
\dot{r}^2 =\frac{E^2}{r^2}\Big[r^2 +\frac{2M}{r}(a-D)^2 -\frac{Q^2}{r^2}(a-D)^2 - \frac{\alpha}{r}ln(\frac{r}{\alpha}) + (a^2 - D^2)\Big]~.
\end{equation}
In general $D\neq a$, but for the trivial case considering $D=a$, we get the geodesic equations to be
\begin{equation}
\frac{dt}{d\lambda}=\frac{r^2 +a^2}{\Delta}E~~;~~\frac{d\phi}{d\lambda}=\frac{a}{\Delta}E~~;~~\frac{dr}{d\lambda}=\pm E~.
\end{equation}
\begin{table}
	\begin{tabular}{|c|c|} 	 	
		\multicolumn{2}{c}{\textbf{ $a$=0.5, $Q$=0.3}}\\
		\hline
		$\alpha$ & $r_{p1}$ \\
		\hline
		0.1 & 2.95\\ 
		\hline
		0.2 &2.755    \\
		\hline
		0.3 & 2.605\\ 
		\hline
		0.4 &2.515    \\
		\hline
		\hline
		0.9 & 2.47\\ 
		\hline
		1.0 &2.505    \\
		\hline
		1.1 & 2.545\\ 
		\hline
		1.2 &2.59    \\
		\hline
	\end{tabular}
	\hspace{2.0cm}
	\begin{tabular}{|c|c|c|} 	 	
		\multicolumn{3}{c}{\textbf{$Q$=0.3, $\alpha$=0.2, 1.0 }}\\
		\hline
		$\alpha$	& $a$ & $r_{p1}$ \\
		\hline
		0.2	&0.1 & 2.39\\ 
		\hline
		0.2	&0.4 &2.67    \\
		\hline
		0.2	&0.7 & 2.92\\ 
		\hline
		0.2 &1.0 &3.15    \\
		\hline
		\hline
		1.0	&0.1 & 2.28\\ 
		\hline
		1.0 &0.4 &2.45    \\
		\hline
		1.0	&0.7 & 2.61\\ 
		\hline
		1.0	&1.0 &2.745\\
		\hline
	\end{tabular}
	\hspace{2.0cm}
	\begin{tabular}{|c|c|c|} 	 	
		\multicolumn{3}{c}{\textbf{$a$=0.5, $\alpha$=0.2, 1.0 }}\\
		\hline
		$\alpha$	& $Q$ & $r_{p1}$ \\
		\hline
		0.2	&0.0 & 2.82\\ 
		\hline
		0.2	&0.3 &2.75    \\
		\hline
		0.2	&0.6 & 2.54\\ 
		\hline
		0.2 &0.9 &2.40    \\
		\hline
		\hline
		1.0	&0.0 & 2.55\\ 
		\hline
		1.0 &0.3 &2.5    \\
		\hline
		1.0	&0.6 & 2.34\\ 
		\hline
		1.0	&0.9 &2.1\\
		\hline
	\end{tabular}
	\caption{\small Radius ($r_{p1}$) of the co-rotating (prograde) photon orbits. }
	\label{35}
\end{table} 
\begin{table}
	\begin{tabular}{|c|c|} 	 	
		\multicolumn{2}{c}{\textbf{ $a$=0.5,$Q$=0.3}}\\
		\hline
		$\alpha$ & $r_{p2}$ \\
		\hline
		0.1 & 1.85\\ 
		\hline
		0.2 &1.69    \\
		\hline
		0.3 & 1.62\\ 
		\hline
		0.4 &1.60    \\
		\hline
		\hline
		0.9 & 1.81\\ 
		\hline
		1.0 &1.87    \\
		\hline
		1.1 & 1.94\\ 
		\hline
		1.2 &2.05    \\
		\hline
	\end{tabular}
	\hspace{2.0cm}
	\begin{tabular}{|c|c|c|} 	 	
		\multicolumn{3}{c}{\textbf{$Q$=0.3, $\alpha$=0.2, 1.0 }}\\
		\hline
		$\alpha$	& $a$ & $r_{p2}$ \\
		\hline
		0.2	&0.1 & 2.18\\ 
		\hline
		0.2	&0.4 &1.83    \\
		\hline
		0.2	&0.7 & 1.345\\ 
		\hline
		\hline
		1.0	&0.1 & 2.15\\ 
		\hline
		1.0 &0.4 &1.95    \\
		\hline
		1.0	&0.7 & 1.7\\ 
		\hline
	\end{tabular}
	\hspace{2.0cm}
	\begin{tabular}{|c|c|c|} 	 	
		\multicolumn{3}{c}{\textbf{$a$=0.5, $\alpha$=0.2, 1.0 }}\\
		\hline
		$\alpha$	& $Q$ & $r_{p2}$ \\
		\hline
		0.2	&0.0 & 1.78\\ 
		\hline
		0.2	&0.3 &1.7    \\
		\hline
		0.2	&0.6 & 1.4\\ 
		\hline
		\hline
		1.0	&0.0 & 1.9\\ 
		\hline
		1.0 &0.3 &1.87    \\
		\hline
		1.0	&0.6 & 1.7\\ 
		\hline
	\end{tabular}
	\caption{\small Radius ($r_{p2}$) of the counter-rotating (retrograde) photon orbits.}
	\label{36}
\end{table} 
For the general case ($D \neq a$), we aim to find the circular photon orbits subject to the conditions $F(r) = F^{\prime}(r)=0$. These two conditions yield 
\begin{equation}\label{2}
r_p ^{2} +\frac{2M}{r_p}(a-D)^2 -\frac{Q^2}{r_p^{2}}(a-D)^2 - \frac{\alpha}{r_p}ln(\frac{r_p}{|\alpha|})(a-D)^2 + (a^2 - D^2)=0
\end{equation}
\begin{equation}\label{1}
2r_p -\frac{2M}{r_p^{2}}(a-D)^2 +\frac{2Q^2}{r_p^{3}}(a-D)^2 + \frac{\alpha}{r_p^{2}}ln(\frac{r_p}{|\alpha|})(a-D)^2-\frac{\alpha}{r_p ^{2}}(a-D)^2=0~.
\end{equation}
Solving for $D$ from eq.\eqref{1}, we get 
\begin{equation}
D=a\mp \sqrt{\frac{2r_p ^{5}}{2Mr_p ^{2}-2Q^2 r_p -\alpha r_p ^{2}ln(\frac{r_p}{|\alpha|})+\alpha r_p ^{2}}}~.
\end{equation}
Here the signs $\mp$ corresponds to the counter rotation (-) and the co-rotation (+) of the orbits along with the black hole. By substituting the expression for $D$ in eq.\eqref{2}, we obtain the following equation which leads to the radius of the photon orbits ($r_p$)
\begin{equation}
6Mr_p ^{3} -4Q^2 r_p ^{2} -3\alpha r_p ^{3}\ln\Big(\frac{r_p}{|\alpha|}\Big)+\alpha r_p ^{3}-2r_p ^{4} \pm 2a\sqrt{2r_p ^{4}\Big(2Mr_p -2Q^2 -\alpha r_{p} \ln\Big( \frac{r_p}{|\alpha|}\Big) +\alpha r_p\Big)}=0~.
\end{equation}
The  $\pm$ signs denote the co-rotating and the counter rotating photon orbits. Tables (\ref{35}) and (\ref{36}) show the photon sphere radius ($r_{p}$) with the variation in spin ($a$), charge ($Q$) and the parameter ($\alpha$) which gives the weightage of the dark matter. Here $r_{p1}$  and $r_{p2}$ correspond to the co-rotating and counter rotating photon sphere radius.

\noindent Since the black hole horizon shows different behaviour in different range of PFDM parameter $\alpha$, so we have analysed the different geodesics and the corresponding characteristics of the particles in two separate ranges. One is the low range of $\alpha$ where $\alpha < \alpha_c$ 
and the other is the range $\alpha >\alpha_c$. 

\begin{itemize}
	
	\item From the above Table(s) (\ref{35}),(\ref{36}), it is quite clear that for the lower range of values of $\alpha < \alpha_c$, the photon radius corresponding to both the co-rotating and the counter-rotating orbits decreases with increase in PFDM parameter.
	
	\item In the higher range values of $\alpha > \alpha_c$, increase in $\alpha$ increases the photon radius as is evident from the nature of the outer event horizon radius ($r_{h+}$).  
	
	\item   With the increase in the value of spin parameter ($a$) of the black hole, the radius of the co-rotating orbits increases while that of the counter rotating orbits decreases for both  $\alpha > \alpha_c$ and $\alpha < \alpha_c$. Since the spin parameter ($a$) of the black hole assists the co-rotation and opposes the counter rotation, thus we observe such feature of photon radius.
	
	\item  The presence of the charge ($Q$) also affects the radius of the photon sphere as can be observed from the Tables (\ref{35}), (\ref{36}). With the increase in the value of charge ($Q$), the photon sphere radius decreases both for prograde and retrograde orbits with $\alpha$=0.2 ($\alpha < \alpha_c$) and $\alpha$=1.0 ($\alpha > \alpha_c$). 
	
	\item Besides we also find that the radius of prograde orbits are larger than the retrograde orbits of the photons moving around the black hole.
	
\end{itemize}
\subsection{Time-like geodesics}
In this case, we consider massive particles and this consideration makes the geodesics time-like, hence we use $\epsilon=-1$. This results in the modification of the radial equation \eqref{100} into the form
\begin{equation}\label{333}
\dot{r}^2 =\Big[E^2 +\frac{2M}{r^{3}}(aE-L)^2 -\frac{Q^2}{r^4}(aE-L)^2 - \frac{\alpha}{r^3}ln\Big(\frac{r}{|\alpha|}\Big) + \frac{1}{r^2}(a^2 E^2 - L^2)\Big]-\frac{\Delta}{r^2}=F(r)~.
\end{equation}
Again for the trivial case $L=aE$, which simplifies the radial equation to the form
\begin{equation}
\frac{dr}{d\tau}=\Big(E^2 -\frac{\Delta}{r^2}\Big)^{\frac{1}{2}}~.
\end{equation}
So the proper time can be evaluated as 
\begin{eqnarray}
\tau = \int \Big(E^2 -\frac{\Delta}{r^2}\Big)^{-\frac{1}{2}}dr~.
\end{eqnarray} 
We now confine ourselves for the general case $(L\neq aE)$. Our aim is to calculate and show the variation of the energy ($E$) and the angular momentum ($L$) of the particle with variation in parameter $\alpha$. So we proceed by assuming $x=L-aE$ and rewrite the above equation in terms of $x$ and  $E$. Upon imposing the conditions for circular orbits ($F(r)=F'(r)=0$), we obtain 
\begin{equation}\label{4}
F(r)=x^2 \Big(a^2 -\Delta\Big) + r^4 E^2 -2aEr^2 x -\Delta r^2=0
\end{equation}
\begin{equation}\label{3}
F'(r)=	4r^3 E^2-4aErx -2\Delta r -\Delta^{'}(r^2 + x^2)=0~.
\end{equation}
Solving the above equations, we obtain the expression for $x$. Using the obtained expression of $x$, we determine the energy ($E$) and angular momentum ($L$) of the particle.

\noindent The expression for $E$ is obtained from eq.(s)\eqref{4}, \eqref{3} as
\begin{equation}\label{Energy}
E=\frac{1}{r^2 ax}\Big[\Big(a^2 -\Delta +\frac{r \Delta^{'}}{4}\Big)x^2 +\Big(\frac{\Delta^{'}}{4}r^3 -\frac{\Delta}{2}r^2\Big)\Big]~.
\end{equation}
On replacing $E$ in eq.\eqref{4} by eq.(\ref{Energy}), we get an equation in $x$ as\footnote{We have shown this in  Appendix B.}	
\begin{align}\label{eq0}
\Big[4\Big(\Delta -a^2- \frac{r \Delta^{'}}{4}\Big)^2 - 4a^2\Big(a^2 -\Delta  +\frac{r \Delta^{'}}{2}\Big)\Big]x^4 + \Big[\Big(4a^2 -4\Delta +r\Delta^{'}\Big) \nonumber \\
\times	\Big(\frac{r^3 \Delta^{'}}{2} -r^2 \Delta\Big)-2r^3 a^2 \Delta^{'}\Big]x^2 + \Big[r^2 \Delta -\frac{r^3 \Delta^{'}}{2}\Big]^2=0~.
\end{align}
The equation is quadratic in $x^2$ with the discriminant
\begin{equation}
\Delta_{D}=16a^2 \Delta^2 r^4 \Big[a^2 - \Delta + \frac{r \Delta^{'}}{2}\Big]~.
\end{equation} 	
Now we can factorize the coefficient of $x^4$ as
\begin{equation}
4\Big(\Delta -a^2- \frac{r \Delta^{'}}{4}\Big)^2 - 4a^2\Big(a^2 -\Delta  +\frac{r \Delta^{'}}{2}\Big)=\mathcal{F_{+}F_{-}}
\end{equation}
where	
\begin{equation}
\mathcal{F_{\pm}}=	2\Big(\Delta -a^2- \frac{r \Delta^{'}}{4}\Big) \pm 2a\sqrt{\Big(a^2 -\Delta  +\frac{r \Delta^{'}}{2}\Big)}~.
\end{equation}
Using the above identities, the expression for $x^2$ can be written down as
\begin{eqnarray}\label{1a}
    x^2 &=& \frac{-\Big[\Big(4a^2 -4\Delta +r\Delta^{'}\Big) \times	\Big(\frac{r^3 \Delta^{'}}{2} -r^2 \Delta\Big)-2r^3 a^2 \Delta^{'}\Big]\pm 4a \Delta r^2\sqrt{ \Big[a^2 - \Delta + \frac{r \Delta^{'}}{2}\Big]}}{2\mathcal{F_{+}}\mathcal{F_{-}}}\nonumber\\
    &=&r^{2}\frac{\Big(\Delta-\mathcal{F_{-}}\Big)}{\mathcal{F_{-}}}~.
\end{eqnarray}
\begin{figure*}[h] 		
	\centering
	\subfigure[$a=0.5$, $Q=0.3$]{\includegraphics[scale=1.2]{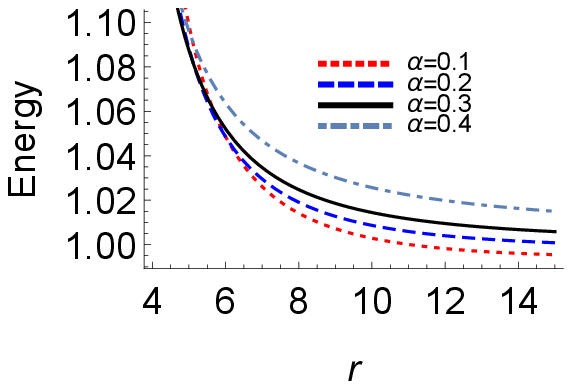}}
	\hfil
	\subfigure[$a=0.5$, $Q=0.3$]{\includegraphics[scale=1.2]{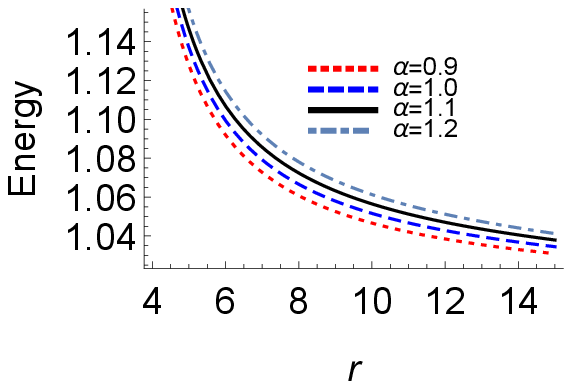}}
	\centering
	\subfigure[$\alpha=0.2$, $Q=0.3$]{\includegraphics[scale=1.2]{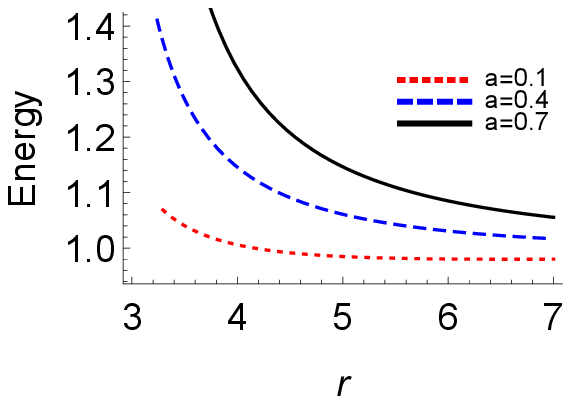}}
	\hfil
	\subfigure[$a=0.5$, $\alpha=0.2$]{\includegraphics[scale=1.2]{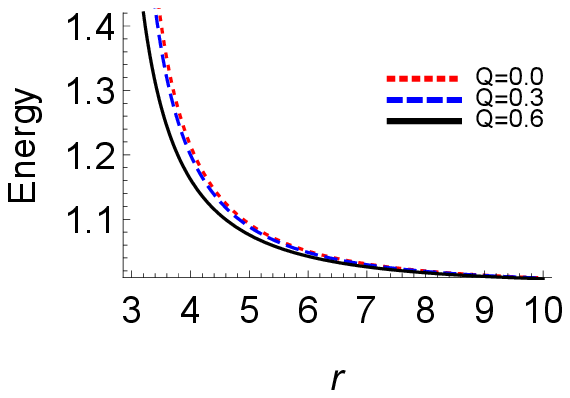}}
	\caption{Plots of energy of the co-rotating massive particles with the variation in $a$, $Q$ and $\alpha$.}
	\label{f1} 
\end{figure*}
The solution for $x$ becomes
\begin{equation}
x=\pm \frac{r}{\sqrt{\mathcal{F_{\mp}}}}\Bigg[a\pm \sqrt{a^2 -\Delta +\frac{r \Delta^{'}}{2}}\Bigg]~.
\end{equation}

\noindent Replacing the values of $x$, the expression for energy becomes	
\begin{equation}\label{E1}
E= \frac{1}{\sqrt{\mathcal{F_{\mp}}}r}\Bigg[\Delta -a\Bigg(a \pm \sqrt{a^2 -\Delta + \frac{r \Delta^{'}}{2}}\Bigg)\Bigg]
\end{equation}	
and that of angular momentum becomes
\begin{equation}
L=\frac{1}{\sqrt{\mathcal{F_{\mp}}}r}\Bigg[a\Bigg(\Delta - a^2 -r^2\Bigg) \mp \Big(r^2 + a^2\Big)\sqrt{a^2 -\Delta + \frac{r \Delta^{'}}{2}}\Bigg]~.
\end{equation}

\noindent The plots of the above expressions for energy ($E$) and angular momentum ($L$) with respect to the radial distance ($r$), gives a firm idea about the orbits and thereby motions around the black hole which in effect gives an impression about the spacetime structure around the black hole. Besides in our case it helps us to get an idea about the impact of the surrounding dark matter on the geodesics. 
\begin{itemize}
	
	\item  In Fig.(\ref{f1}), we have graphically represented the  energy of a co-rotating particle, given in eq.(\ref{E1}). We find that the energy ($E$) of the co-rotating particle falls with distance ($r$) from the black hole also in all cases the energy approaches unity which is the energy of a particle observed by a stationary observer at infinity.
	
	\item From Fig.(\ref{f1}) it can be observed that while the co-rotating particle is near the black hole, it is assisted by the spin ($a$) of the black hole and its energy thereby increases but as it starts to move away, its energy decreases.
\end{itemize}
\begin{figure*}[h]
	\centering
	\subfigure[$a=0.5$, $Q=0.3$]{\includegraphics[scale=1.2]{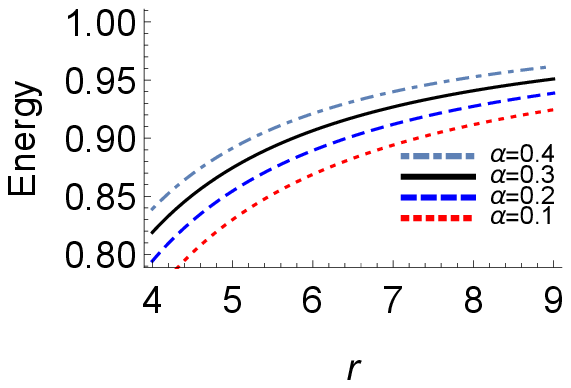}}
	\hfil
	\subfigure[$a=0.5$, $Q=0.3$]{\includegraphics[scale=1.2]{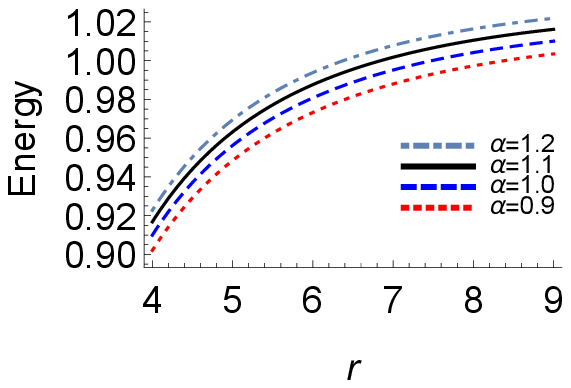}}
	\caption{Plots of energy of the counter-rotating massive particles with variation in $\alpha$.}
	\label{f2}
\end{figure*} 

\begin{itemize}
	
	\item  The effect of the dark matter ($\alpha$) is similar for both lower values ($\alpha < \alpha_c$) and higher values ($\alpha > \alpha_c$) as evident from Fig.(\ref{f1}). In both cases, energy increases with increase in the value of the PFDM parameter $\alpha$.
	
	\item  The increase in spin ($a$) for a fixed value of $\alpha$ (which in this case is taken to be $\alpha=0.2$)  results in the increment of the energy of the particle which means an assist by the rotation of the black hole. This has been shown in Fig.(\ref{f1}). 
	
	\item On the other hand for a fixed value of spin $a$ and PFDM parameter $\alpha$, increase in charge ($Q$) results in decrease in the energy of the particle as can be seen from Fig.(\ref{f1}). A possible reason may be that for a fixed spin, the mass of the black hole increases with increase in the charge and hence the energy of the black hole increases which results in a decrease in the rotational energy of the particles.
\end{itemize}

\begin{wrapfigure}[15]{r}{0.5\textwidth}
	\begin{center}
		\includegraphics[width=0.4\textwidth]{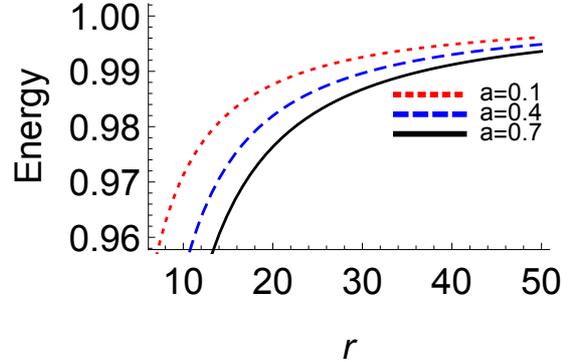}
	\end{center}
	\caption{\small Plots of energy of the counter-rotating massive particles for $\alpha$=0.2 and $Q$=0.3 with spin ($a$) variation.}
	\label{f100}
\end{wrapfigure}
\noindent In Fig.(\ref{f2}), we find that energy ($E$) of the counter-rotating particle increases with distance ($r$) from the black hole and as it moves away it slowly approaches unity. As the particle is closer it is opposed by the black hole rotation and hence it has less energy.
\begin{itemize}
	\item In this case, the effect of dark matter ($\alpha$) is similar for both low values ($\alpha < \alpha_c$) and high values ($\alpha > \alpha_c$). In both cases (varying $\alpha$) energy increases with increase in the value of PFDM parameter $\alpha$. This has been shown in Fig.(\ref{f2}). 
	
	\item The increase in spin ($a$) of the black hole for a fixed value of $\alpha$ results in decrement of energy of the particle as is evident from Fig.(\ref{f100}). A possible reason may be that with increase in spin of the black hole, the rotational kinetic energy of the black hole increases resulting in the decrease in the energy of the particles in order to keep the total energy fixed.
\end{itemize}

\begin{figure*}[h]
	\centering
	\subfigure[$a=0.5$, $Q=0.3$]{\includegraphics[scale=1.2]{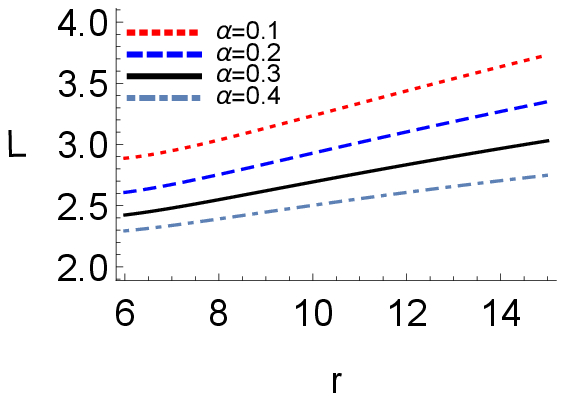}}
	\hfil
	\subfigure[$a=0.5$, $Q=0.3$]{\includegraphics[scale=1.2]{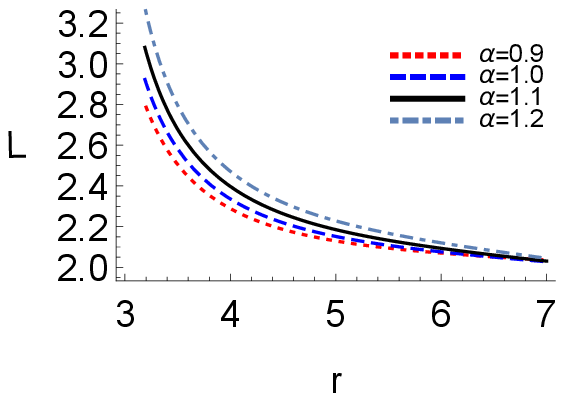}}
	\caption{Plots of angular momentum of the massive particles co-rotating with the black hole with variation in $\alpha$ with fixed spin($a$) and charge($Q$).}
	\label{f3}
\end{figure*}

\noindent The plots of the angular momentum present a completely different picture with respect to the different range (higher and lower values) of the PFDM parameter $\alpha$. 
\begin{itemize}
	
	\item  Fig.(\ref{f3}) shows the angular momentum of the particles moving in prograde (co-rotating) orbits. The plots show that in the lower range of $\alpha$ ($\alpha < \alpha_c$), increasing the value of $\alpha$ decreases the angular momentum whereas the reverse is observed in case of the higher range of  $\alpha$ ($\alpha > \alpha_c$).  
\end{itemize}

\begin{figure*}[h]
	\centering
	\subfigure[$a=0.5$, $Q=0.3$]{\includegraphics[scale=1.2]{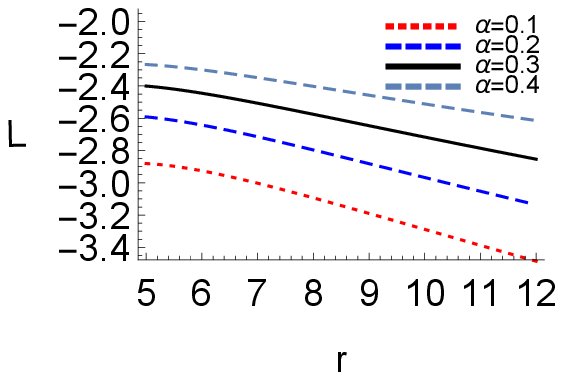}}
	\hfil
	\subfigure[$a=0.5$, $Q=0.3$]{\includegraphics[scale=1.2]{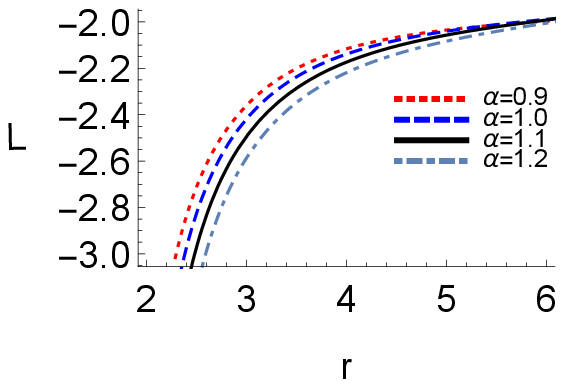}}
	\caption{Plots of angular momentum of the massive particles counter-rotating with the black hole with variation in $\alpha$ with fixed values of spin($a$) and charge($Q$).}
	\label{f4}
\end{figure*} 
\begin{itemize}
	\item   The angular momentum of the particles in the retrograde (counter-rotating) orbits are negative since they move opposite to the direction of rotation of the black hole as shown in Fig.(\ref{f4}). It can be observed that for $\alpha < \alpha_c$,  increase in the value of $\alpha$ reduces the magnitude of the angular momentum of the massive test particle. However, for $\alpha > \alpha_c$, the magnitude of the angular momentum gets increased. 
	
\end{itemize}
\subsection{Geodesics of charged particle}
After studying the null-geodesics and geodesics of the chargeless massive particles, we move on to study the geodesic motion of the massive charged particles. Incorporating the interactions of the gauge fields, the Hamiltonian of the particle in this case gets modified to \cite{28} 
\begin{equation}
2\mathcal{H}=g^{\mu \nu}\Big(p_{\mu} + qA_{\mu}\Big)\Big(p_{\nu} + qA_{\nu}\Big)=\epsilon=-1
\end{equation}
with the electromagnetic potential for charged spinning black hole coupled to PFDM given by \cite{33} 
\begin{equation}
A=A_{\mu}dx^{\mu}=\frac{Qr}{\rho^2}\Big(dt-a\sin^2 \theta d\phi\Big)~.
\end{equation}
Using the Legendre transformation $\mathcal{H}=p_{\mu}\dot{x}^{\mu}-\mathcal{L}$, we obtain the Lagrangian of the particle as
\begin{equation}
\mathcal{L}=\frac{1}{2}g_{\mu \nu}\dot{x}^{\mu}\dot{x}^{\nu}-qA_{\mu}\dot{x}^{\mu}
\end{equation}
where $q$ denotes the charge of the particle. Also we consider the particle to have unit mass ($m=1$) for the sake of simplicity. Using the symmetry of the metric we compute the conserved quantities $E$ and $L$ which have same physical meaning as mentioned previously. Since we are interested in the equatorial geodesics, we use $\theta = \frac{\pi}{2}$ and $\dot{\theta}=0$ which leads to following the geodesic equations 
\begin{equation}
\dot{t}=\frac{1}{r^2}\Bigg[\frac{r^2 +a^2}{\Delta}\Bigg(\Big(E-\frac{qQ}{r}\Big)\Big(r^2 +a^2\Big)-a\Big(L-\frac{qaQ}{r}\Big)\Bigg)+a\Bigg(\Big(L-\frac{qaQ}{r}\Big)-a\Big(E-\frac{qQ}{r}\Big)\Bigg)\Bigg]
\end{equation}
\begin{figure*}[h]
	\centering
	{\includegraphics[scale=1.2]{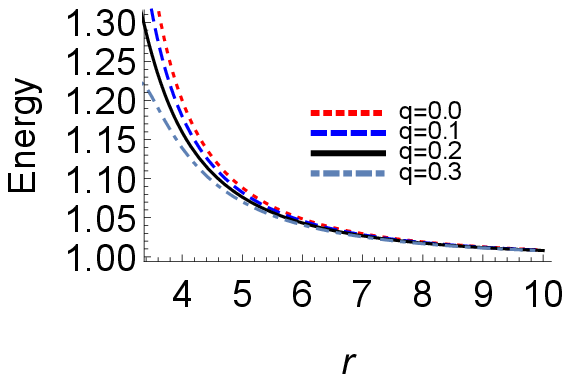}}
	\hfil
	{\includegraphics[scale=1.2]{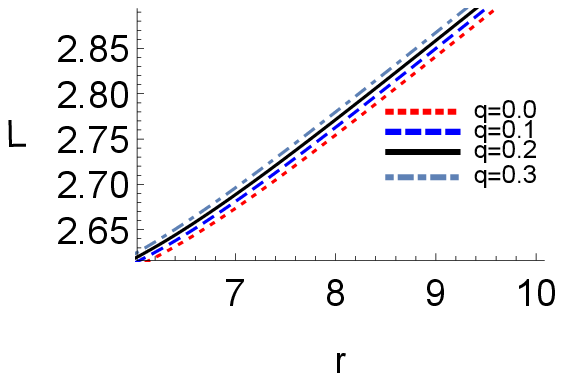}}
	\caption{Plots for energy and angular momentum of charged particles co-rotating with the black hole with varying $q$ for $\alpha$=0.2, $Q$=0.3 and $a$=0.5.}
	\label{f5}
\end{figure*}
\begin{figure*}[h]
	\centering
	{\includegraphics[scale=1.2]{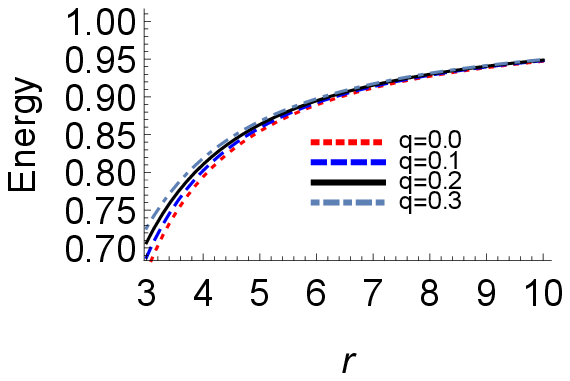}}
	\hfil
	{\includegraphics[scale=1.2]{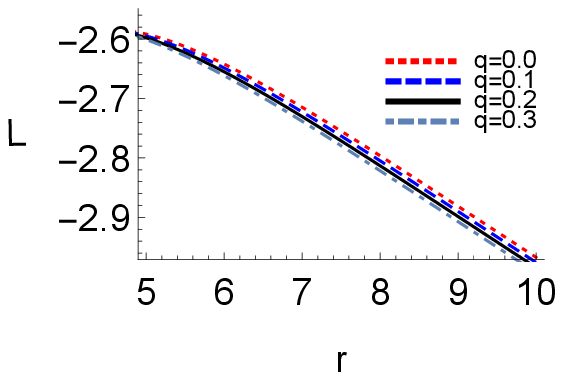}}
	\caption{Plots for energy and angular momentum of charged particles counter-rotating with the black hole with varying $q$ for $\alpha$=0.2, $Q$=0.3 and $a$=0.5.}
	\label{f6}
\end{figure*}
\begin{equation}
\dot{\phi}=\frac{1}{r^2}\Bigg[\frac{a}{\Delta}\Bigg(\Big(E-\frac{qQ}{r}\Big)(r^2 +a^2)-a\Big(L-\frac{qaQ}{r}\Big)\Bigg)+\Bigg(\Big(L-\frac{qaQ}{r}\Big)-a\Big(E-\frac{qQ}{r}\Big)\Bigg)\Bigg]
\end{equation}
\begin{align}\label{12}
\dot{r}^2 = -\frac{\Delta}{r^2} + \frac{\Big(E-\frac{qQ}{r}\Big)^2}{r^4}\Bigg[\Big(r^2 + a^2\Big)^2 -a^2 \Delta\Bigg]-\frac{2a}{r^4}\Big(r^2 + a^2 - \Delta \Big)\Big(E-\frac{qQ}{r}\Big)\Big(L-\frac{qaQ}{r}\Big)  \nonumber \\
-\frac{1}{r^4}\Big(\Delta -a^2\Big)\Big(L-\frac{qaQ}{r}\Big)^2~.
\end{align}
In order to determine the circular orbits, we use the conditions $F(r)=F'(r)=0$ where 
\begin{align}
F(r)=-\frac{\Delta}{r^2} + \frac{\Big(E-\frac{qQ}{r}\Big)^2}{r^4}\Bigg[\Big(r^2 + a^2\Big)^2 -a^2 \Delta\Bigg]-\frac{2a}{r^4}\Big(r^2 + a^2 - \Delta \Big)\Big(E-\frac{qQ}{r}\Big)\Big(L-\frac{qaQ}{r}\Big)  \nonumber \\
-\frac{1}{r^4}\Big(\Delta -a^2\Big)\Big(L-\frac{qaQ}{r}\Big)^2~.
\end{align}
\begin{figure*}[h]
	\centering
	{\includegraphics[scale=1.2]{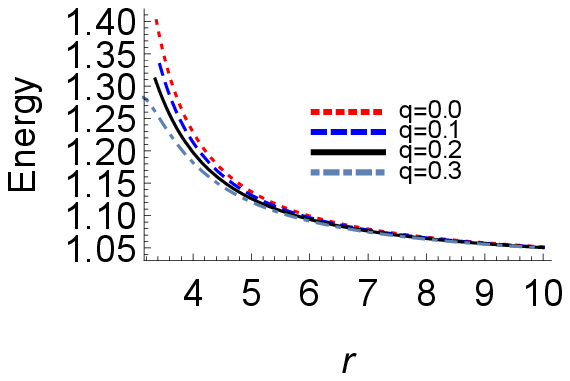}}
	\hfil
	{\includegraphics[scale=1.2]{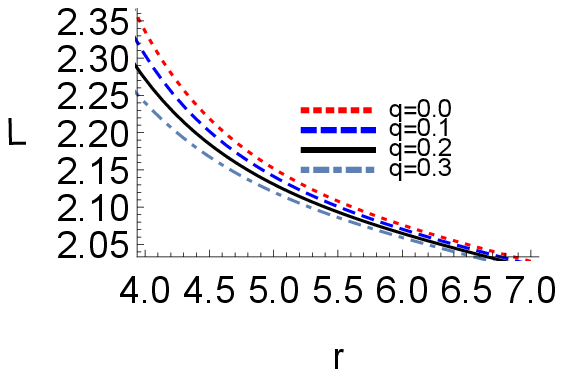}}
	\caption{Plots for energy and angular momentum of charged particles co-rotating with the black hole with varying $q$ for $\alpha$=1.0, $Q$=0.3 and $a$=0.5.}
	\label{f7}
\end{figure*}
\begin{figure*}[h]
	\centering
	{\includegraphics[scale=1.2]{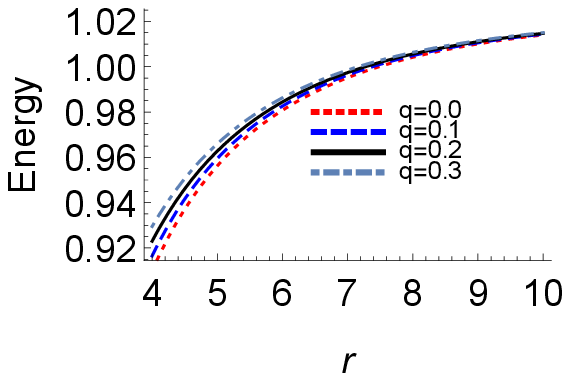}}
	\hfil
	{\includegraphics[scale=1.2]{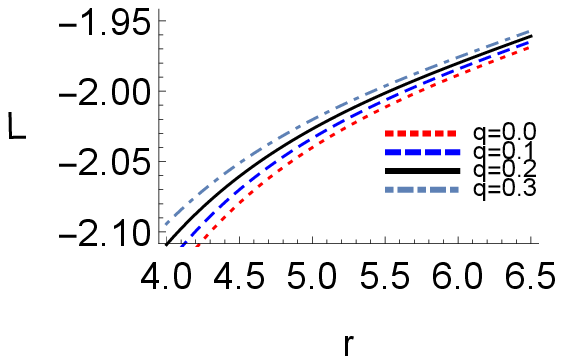}}
	\caption{Plots for energy and angular momentum of charged particles counter-rotating with the black hole with varying $q$ for $\alpha$=1.0, $Q$=0.3 and $a$=0.5.}
	\label{f8}
\end{figure*}  

\noindent  It is very difficult to find an exact solution for $E$ and $L$ from these conditions. So we proceed by assuming $\frac{q}{m} \ll 1$ (particle with small specific charge) and approximately write down the following solutions by incorporating Taylor expansion about $\frac{q}{m} = 0$ \cite{26}
\begin{equation}
E(q)=E(0) + qE'(0) + O(q^2)+...
\end{equation}
\begin{equation}
L(q)=L(0) + qL'(0) + O(q^2) +...~.
\end{equation}
The approximate solutions of $E$ and $L$ satisfy the condition of circular orbits given as $F(r)=F'(r)=0$. We display the plots. 
\begin{itemize}
	\item From Fig(s).(\ref{f5}), (\ref{f6}), (\ref{f7}) and (\ref{f8}), it can be observed that for both $\alpha < \alpha_c$ and $\alpha > \alpha_c$,  the increase in the value of the charge of the particle $q$ from  $0.0$ to $0.3$, decreases the energy of co-rotating particles whereas there is an increase in the energy for counter rotating particles. In both cases the energy tends towards unity. 
	
	\item  In case of angular momentum, we observe that for $\alpha=0.2$ the angular momentum of both co-rotating ($+$ve increase) and counter rotating ($-$ve increase) particles  increases with the increase in the value of the charge $q$ of the particle. However, for $\alpha=1.0$ we find that with increase in the value of charge $q$, angular momentum ($L$) for co-rotating and counter rotating particles decreases.
	
	\item The observations depict that the particle's charge $q$ responses differently depending on the value of PFDM parameter $\alpha$. Also it implies that the particle with more charge ($q$) is hindered more if the intensity of dark matter increases.
	
\end{itemize}
\begin{figure*}
	\centering
	\subfigure[$a=0.5$, $Q=0.3$, $L=3.0$]{\includegraphics[scale=1.2]{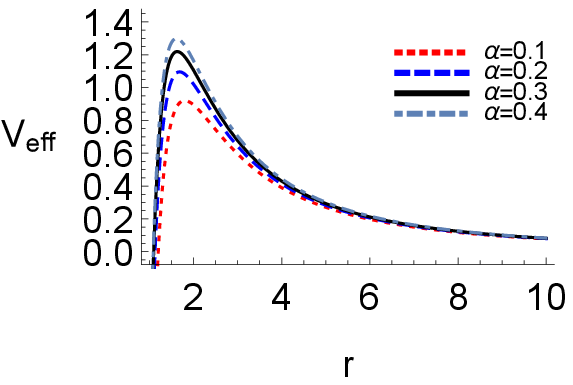}}
	\hfil
	\subfigure[$a=0.5$, $Q=0.3$, $L=3.0$]{\includegraphics[scale=1.2]{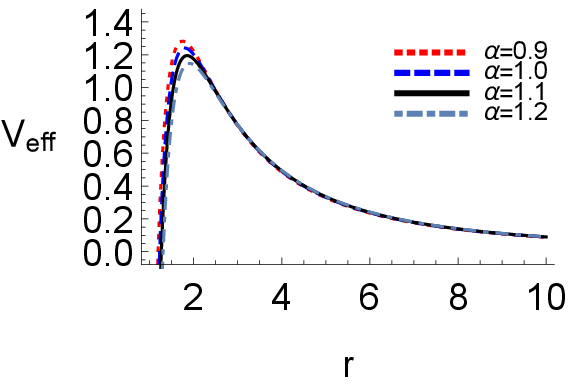}}
	\centering
	\subfigure[$\alpha$=0.2, $Q=0.3$, $L=3.0$]{\includegraphics[scale=1.2]{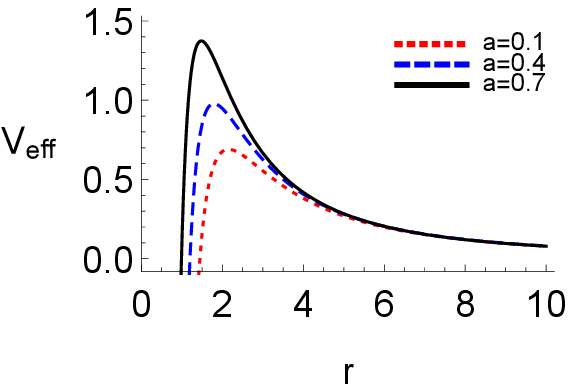}}
	\hfil
	\subfigure[$a=0.5$, $\alpha$=0.2, $L=3.0$]{\includegraphics[scale=1.2]{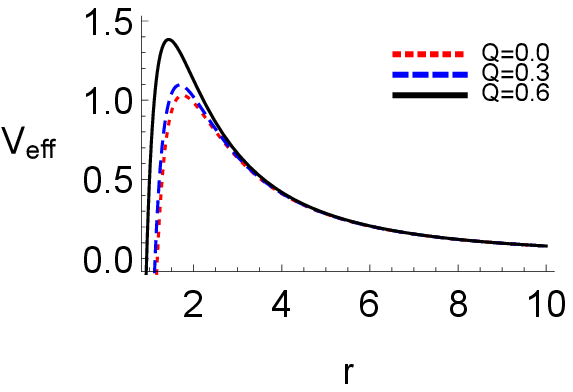}}
	\centering
	\subfigure[$a=0.5$, $\alpha$=0.2, $Q=0.3$]{\includegraphics[scale=1.2]{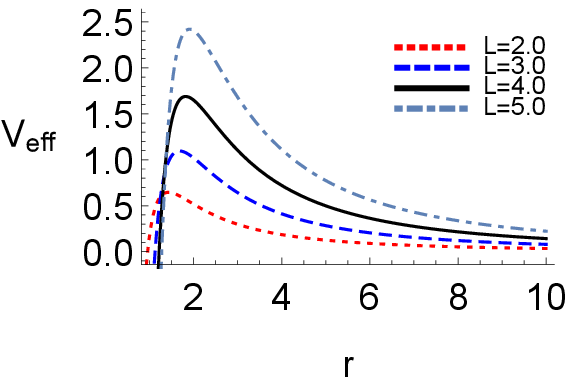}}
	\hfil
	\subfigure[$a=0.5$, $\alpha$=1.0, $Q=0.3$]{\includegraphics[scale=1.2]{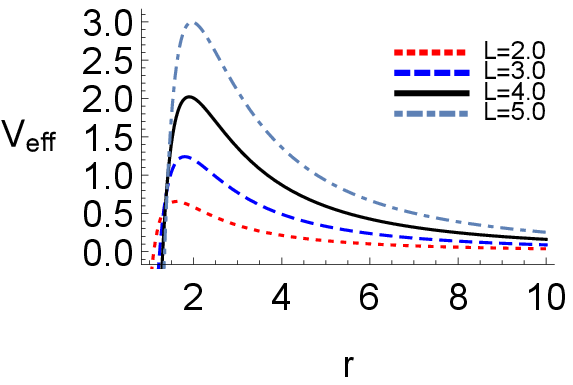}}
	\caption{Plots of effective potential for null geodesics with variation in $a$, $Q$, $L$ and $\alpha$.}
	\label{f9}
\end{figure*}
\begin{figure*}
	\centering
	\subfigure[$a=0.5$, $Q=0.3$, $L=3.0$]{\includegraphics[scale=1.2]{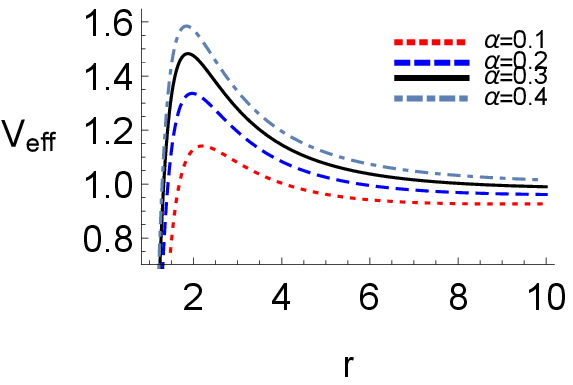}}
	\hfil
	\subfigure[$a=0.5$, $Q=0.3$, $L=3.0$]{\includegraphics[scale=1.2]{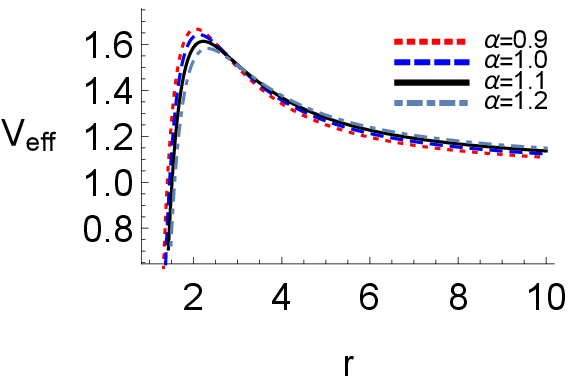}}
	\centering
	\subfigure[$\alpha$=0.2, $Q=0.3$, $L=3.0$]{\includegraphics[scale=1.2]{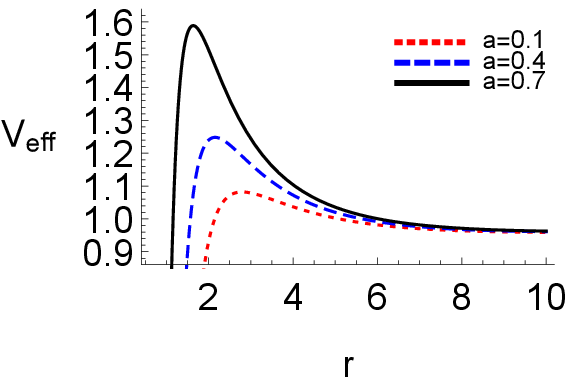}}
	\hfil
	\subfigure[$a=0.5$, $\alpha$=0.2, $L=3.0$]{\includegraphics[scale=1.2]{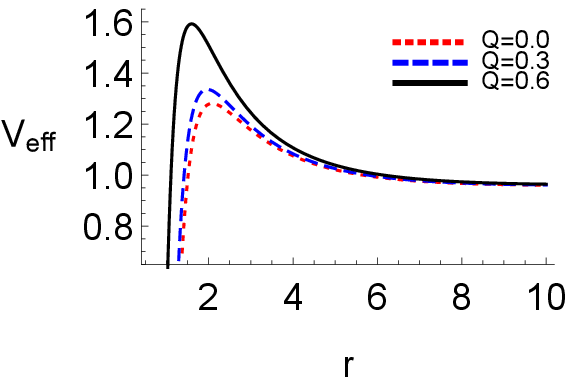}}
	\centering
	\subfigure[$a=0.5$, $\alpha$=0.2, $Q=0.3$]{\includegraphics[scale=1.2]{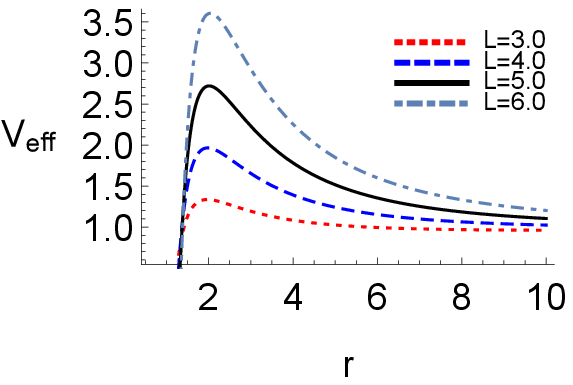}}
	\hfil
	\subfigure[$a=0.5$, $\alpha$=1.0, $Q=0.3$]{\includegraphics[scale=1.2]{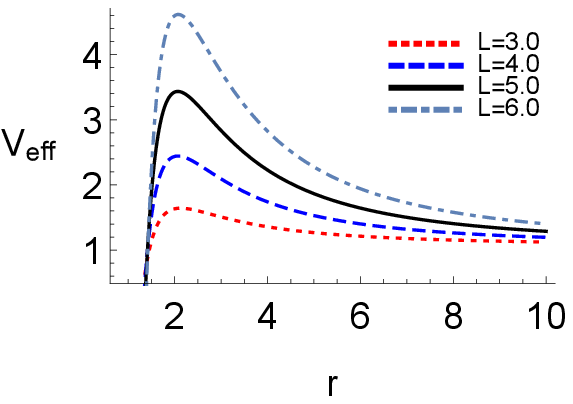}}
	\caption{Plots of effective potential for timelike geodesics with variation in $a$, $Q$, $L$ and $\alpha$.}
	\label{f10}
\end{figure*}

\section{Nature of the effective potential}\label{sec5}
In this section we study the effective potential ($V_{eff}$) which results in both stable and unstable orbits depending upon the condition $ \frac{\partial ^2 V_{eff}}{\partial r^2} > 0$	or $ \frac{\partial ^2 V_{eff}}{\partial r^2} < 0$ respectively. The stable and unstable orbits correspond to the local minima and maxima of the potential which we obtain from the radial geodesic equations. The potential depends upon the following parameters, the charge of the black hole ($Q$), spin parameter ($a$), the dark matter parameter $\alpha$ and on the charge of the particle ($q$). The effective potential in case of the massive particles obtained from the corresponding radial geodesic equation eq.\eqref{333} reads 
\begin{equation}\label{10}
\dot{r}^2 + V_{eff} =E^2
\end{equation}
where the effective potential $V_{eff}$ is given by
\begin{equation} 
V_{eff}=-\frac{2M}{r^{3}}(aE-L)^2 +\frac{Q^2}{r^4}(aE-L)^2 + \frac{\alpha}{r^3}\ln\Big(\frac{r}{|\alpha|}\Big) - \frac{1}{r^2}(a^2 E^2 - L^2) + \frac{\Delta}{r^2}~. 
\end{equation}
For circular geodesics, the particle moves in a circular trajectory of constant radius $r$ which implies $\dot{r}=0$. In case of photons, the effective potential as obtained from eq.\eqref{363} takes the form 
\vspace{-0.1cm}
\begin{equation}
V_{eff}=-\frac{2M}{r^{3}}(aE-L)^2 +\frac{Q^2}{r^4}(aE-L)^2 + \frac{\alpha}{r^3}\ln\Big(\frac{r}{|\alpha|}\Big) - \frac{1}{r^2}(a^2 E^2 - L^2)~.
\end{equation}
Also for massive particles with charge $q$, the effective potential takes the following form obtained using eq.\eqref{12} 
\begin{align}
V_{eff}=E^2 + \frac{\left(\Delta-a^2\right) \left(L-\frac{a q Q}{r}\right)^2}{r^4}+\frac{2a \left(a^2-\Delta +r^2\right) \left(E-\frac{q Q}{r}\right) \left(L-\frac{a q Q}{r}\right)}{r^4} \nonumber \\ -\frac{\left(\left(a^2+r^2\right)^2-a^2 \Delta\right) \left(E-\frac{q Q}{r}\right)}{r^4}+\frac{\Delta}{r^2}~.
\end{align}
In the trivial case when $L=aE$, the radial equations and effective potentials become
\begin{equation}
\dot{r}_{massive} = \pm \sqrt{E^2 -\frac{\Delta}{r^2}}~~;~~ \dot{r}_{null}= \pm E
\end{equation}
\begin{equation}
V_{massive}=\frac{\Delta}{r^2}~~;~~V_{null}=0~.
\end{equation} 
\noindent In Fig.(\ref{f9}), we show the plots of the effective potential for null geodesics and the plots corresponding to the effective potential for timelike geodesics are given in Fig.(\ref{f10}).  
\begin{itemize}
	
	\item From Fig.(s)(\ref{f9}) and (\ref{f10}) we find that the effective potential of both null and timelike geodesics,  increases with increase in the value of PFDM parameter $\alpha$ for $\alpha <\alpha_c$. However for $\alpha > \alpha_c$ the effective potential falls with increase in the PFDM parameter $\alpha$.
	
	\item The potential increases with increase in the spin ($a$) and charge ($Q$) of the black hole both for $\alpha < \alpha_c$ and $\alpha > \alpha_c$. This has been observed for effective potential corresponding to both null and timelike geodesics. 
	
	\item Besides we observe that in both cases, with increase in angular momentum ($L$) of the particle, the effective potential rises in both cases ($\alpha < \alpha_c$ and $\alpha > \alpha_c$) and the maxima shifts towards larger radial distance $r$.
\end{itemize}

\section{Penrose process}\label{sec6}
Black hole is a vessel of extreme energy and there are many processes theorised which are
responsible to gain energy (extract energy) from the black hole. One of them is the Penrose process named
after Roger Penrose who proposed the mechanism in \cite{44}. In case of rotating black hole, a region gets created between the outer event horizon ($g^{rr}=0$) and the stationary limit surface ($g_{tt}=0$). These two surfaces meet at the poles and have largest separation in the equatorial plane. This varying annular region is known as the \textit{ergosphere}. In case of static black hole this region vanishes. The speciality of this region is that the Killing vector $\frac{\partial}{\partial t}$ which has a unit norm as observed by a stationary observer at infinity becomes spacelike within the region. The symmetry of the metric with change in the said Killing vector results in energy conservation and hence the energy in this region can be negative. This fact can be utilised to gain energy (extract energy) from the black hole.  
\subsection{Extraction of energy from the black hole}
\noindent Let a particle (uncharged) with positive energy fall into this ergoregion and split into two particles, one with positive energy and the other with negative energy. The negative energy particle is absorbed by the black hole and that with positive energy comes out of the black hole having more energy than the particle that entered the black hole and hence resulting in energy gain (effective energy extraction).

\noindent The condition of negative energy of the particle can be found using the condition of circular orbits. The equation with $\dot{r}=0$ results in
\begin{equation}
E^2\Big[\Big(r^2 + a^2\Big)^2 -a^2 \Delta\Big] -E\Big[2aL\Big(r^2 + a^2 -\Delta\Big) + L^2 \Big(a^2 -\Delta\Big) + \Delta r^2 \epsilon\Big]=0  
\end{equation}	
which can be solved for both $E$ and $L$ as given by
\begin{equation}\label{7}
E=\frac{aL\Big(r^2 + a^2 -\Delta\Big)\pm r\sqrt{\Delta \Big[r^2 L^2 -\epsilon\Big(\Big(r^2 + a^2\Big)^2 -a^2 \Delta\Big)\Big]}}{\Big(r^2 + a^2\Big)^2 -a^2 \Delta}
\end{equation}	
\begin{equation}
L=\frac{aE\Big(r^2 + a^2 -\Delta\Big)\pm r\sqrt{\Delta \Big[r^2 E^2 +\epsilon\Big( \Delta - a^2\Big)\Big]}}{\Big(a^2 -\Delta\Big)}~.
\end{equation}		
If one assumes positive sign in eq.(\ref{7}) along with the condition
\begin{equation}
a^2 L^2\Big(r^2 + a^2 -\Delta\Big)^2 > \Delta r^2 \Big[r^2 L^2 -\epsilon\Big(\Big(r^2 + a^2\Big)^2 -a^2 \Delta\Big)\Big]
\end{equation}	
and $L<0$, then $E<0$, i.e., particle with negative energy is possible. This gives the idea that $E<0$ is possible for $L<0$ which is the case for counter rotating orbits. The negative energy particles following counter rotating orbits must lie within the ergoregion. The expression for the negative energy particles take the form
\begin{equation}
  E=\frac{aL\Big(r^2 + a^2 -\Delta\Big)-r\sqrt{\Delta \Big[r^2 L^2 -\epsilon\Big(\Big(r^2 + a^2\Big)^2 -a^2 \Delta\Big)\Big]}}{\Big(r^2 + a^2\Big)^2 -a^2 \Delta}~.  
\end{equation}
The plots of negative energy with the variation in different parameters are shown below.

\noindent In order to discuss the Penrose process in detail, we must start by considering an uncharged particle of energy $E_{0}$ entering the ergosphere and let, it breaks down into two photons with energies $E_{1}$ and $E_{2}$. Let the angular momentum of the particles be $L_{0}$ (entering), $L_{1}$ (leaving) and $L_{2}$ (captured). Also let the energy of the particle entering the ergosphere be $E_{0}=1$. Hence the angular momentum of the particles are
\begin{equation}
L_{0}=\frac{a\Big(r^2 + a^2 -\Delta\Big) + r\sqrt{\Delta \Big[r^2  - \Big( \Delta - a^2\Big)\Big]}}{\Big(a^2 -\Delta\Big)}~~;~~\epsilon=-1
\end{equation}	
\begin{equation}
L_{1}=\frac{aE_{1}\Big(r^2 + a^2 -\Delta\Big) + r\sqrt{\Delta \Big(r^2E_{1}^2 \Big)}}{\Big(a^2 -\Delta\Big)}=b_{1}E_{1}
\end{equation}	
\begin{equation}
L_{2}=\frac{aE_{2}\Big(r^2 + a^2 -\Delta\Big) - r\sqrt{\Delta \Big(r^2E_{2}^2 \Big)}}{\Big(a^2 -\Delta\Big)}=b_{2}E_{2}
\end{equation}
\begin{figure*}
	\centering
	\subfigure[$a=0.5$, $Q=0.3$, $L=-3.0$]{\includegraphics[scale=1.2]{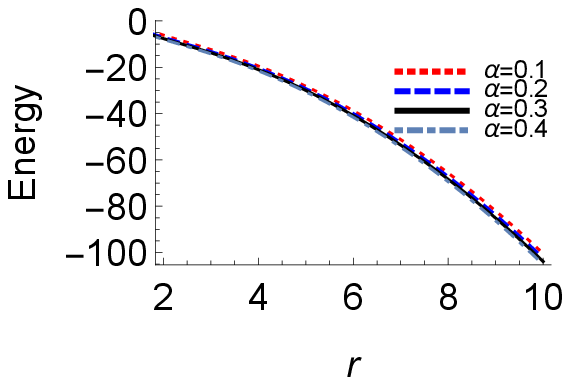}}
	\hfil
	\subfigure[$a=0.5$, $Q=0.3$, $L=-3.0$]{\includegraphics[scale=1.2]{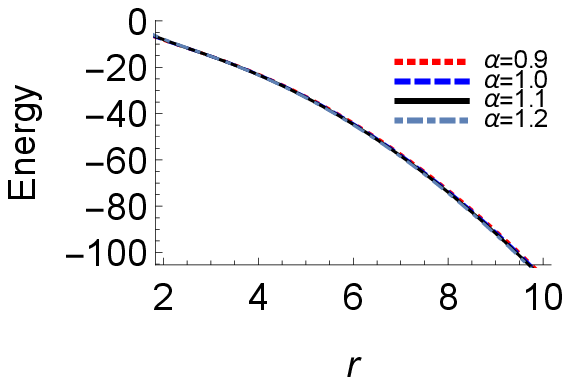}}
	\centering
	\subfigure[$a=0.5$, $\alpha$=0.2, $Q=0.3$]{\includegraphics[scale=1.2]{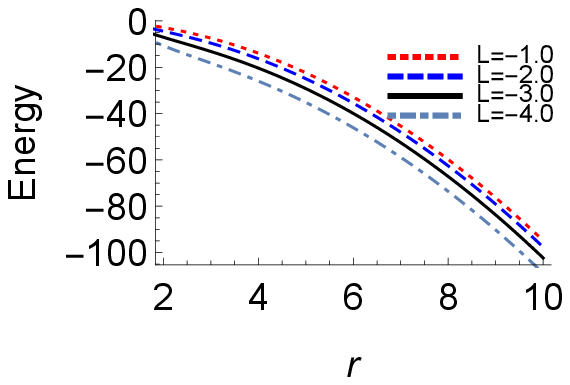}}
	\hfil
	\subfigure[$a=0.5$, $\alpha$=1.0, $Q=0.3$]{\includegraphics[scale=1.2]{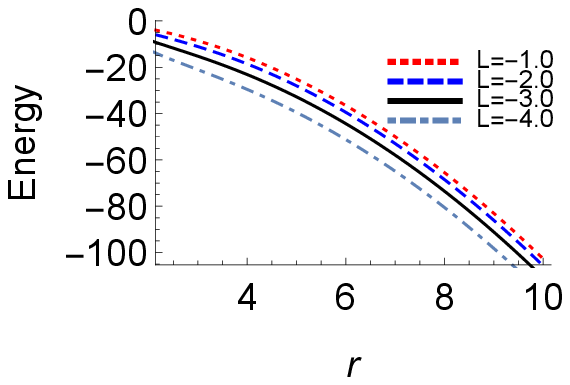}}
	\caption{Plots for negative energy particle with variation in $L$ and $\alpha$.}
	\label{f13}
\end{figure*} 
\begin{figure*}
	\centering
	\subfigure[$a=0.5$, $Q=0.3$]{\includegraphics[scale=1.2]{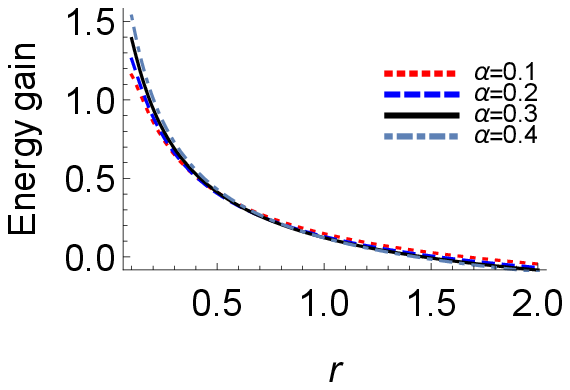}}
	\hfil
	\subfigure[$a=0.5$, $Q=0.3$]{\includegraphics[scale=1.2]{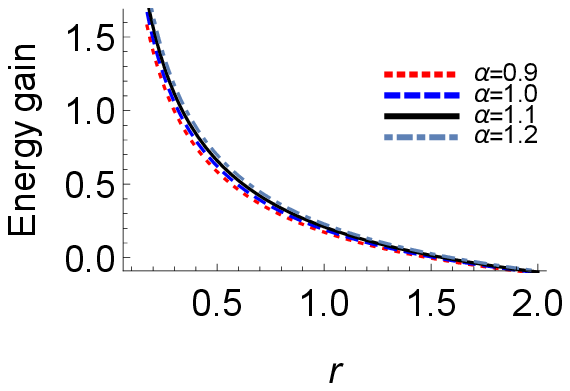}}
	\centering
	\subfigure[$\alpha$=0.2, $Q=0.3$]{\includegraphics[scale=1.2]{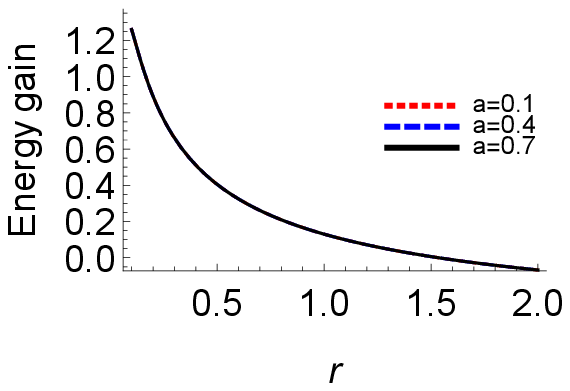}}
	\hfil
	\subfigure[$a=0.5$, $\alpha$=0.2]{\includegraphics[scale=1.2]{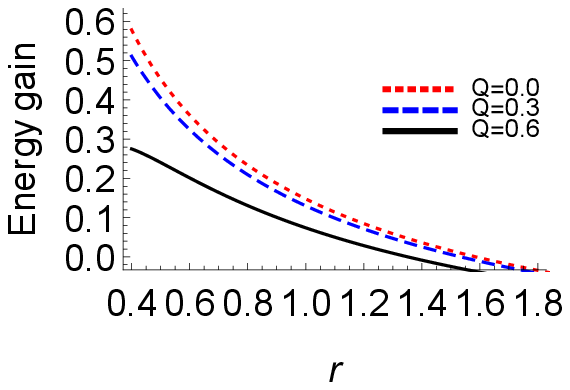}}
	\caption{Plots showing energy gain from the black hole with variation in $a$, $Q$ and $\alpha$.}
	\label{f14}
\end{figure*}		
where 
\begin{equation}
b_{1}=\frac{a\Big(r^2 + a^2 -\Delta\Big) +  r^2\sqrt{\Delta}}{\Big(a^2 -\Delta\Big)} ~~;~~b_{2}=\frac{a\Big(r^2 + a^2 -\Delta\Big) -  r^2\sqrt{\Delta}}{\Big(a^2 -\Delta\Big)}~.
\end{equation}	
By conservation of energy and angular momentum we get
\begin{equation}
E_{0}=E_{1}+E_{2}=1~~;~~L_{0}=b_{1}E_{1}+b_{2}E_{2}~.
\end{equation}	
Solving for $E_{1}$ and $E_{2}$ we obtain 
\begin{equation}
E_{1}=\frac{1}{2}\Bigg[1+\sqrt{\frac{r^2 +a^2 -\Delta}{r^2}}\Bigg]
\end{equation}	
\begin{equation}
E_{2}=\frac{1}{2}\Bigg[1-\sqrt{\frac{r^2 + a^2 -\Delta}{r^2}}\Bigg]
\end{equation}	  
where $E_{1}$ and $E_{2}$ corresponds to the positive and negative energies of the two particles. Thus the particle with energy $E_{2}$ is captured by the black hole while that with energy $E_{1}$ comes out of the black hole resulting in an energy gain of 
\begin{eqnarray}\label{11}
\Delta E =E_{1}-1 &=& -E_{2}\nonumber\\
 &=& \frac{1}{2}\Bigg[\sqrt{\frac{r^2 + a^2 -\Delta}{r^2}}-1 \Bigg]\nonumber\\
  &=& \frac{1}{2}\Bigg[\sqrt{\frac{2Mr - Q^2 - \alpha r \ln\Big(\frac{r}{|\alpha|}\Big)}{r^2}}-1 \Bigg]~.   
\end{eqnarray}
In the limit $a \to 0$, the ergosphere vanishes and the region of ergosphere corresponds to event horizon with radius $r_{h+}$ and hence $\Delta =0$ and we get energy gain $\Delta E =0$, $E_{1}=1$ and $E_{2}=0$ and hence no particle with negative energy exists.
In the limit $Q\rightarrow 0$ and $\alpha \rightarrow 0$, eq.(\ref{11}) matches with the result given in \cite{2}.
\noindent The plots of negative energy and energy gain from the black hole are shown above. The plots depict how the negative energy states depend on the parameters characterising the black hole spacetime.
\begin{itemize}
	
	\item  From Fig.(\ref{f13}), we find that for $\alpha < \alpha_c$, the negative energy increases with increase in $\alpha$ and similar is true for $\alpha > \alpha_c$. However the change is less prominent for higher values. 
	
	\item The influence of charge ($Q$) and angular momentum ($L$) are firmly observed, where for both large and small constant values of $\alpha$, the increase in the charge and the negative angular momentum increases the negative energy of the particle quite impressively which results in the fact that the particle absorbed by the black hole will have higher negative energy and will lead to increased energy gain from the black hole. 
\end{itemize}

\noindent The energy gain from the black hole via., the Penrose process is astrophysically very important and significant. Also indirectly more the negative energy absorbed by the black hole, more is the gain in positive energy via Penrose process. The plots of energy gain in  Fig.(\ref{f14}) show us the impact of different black hole parameters on the proportion of increment or decrement of the gain in the energy.
\begin{itemize}
	
	\item  We find that energy gain increases with increase in the PFDM parameter $\alpha$ for both $\alpha < \alpha_c$ and $\alpha > \alpha_c$. 
	
	\item  With increase in the value of the charge ($Q$), energy gain decreases for $\alpha < \alpha_c$. The expression for energy gain given eq.(\ref{11}) is independent of the spin parameter $a$. This implies that variation in the spin parameter $a$ will not effect the energy extraction. This can be observed in Fig.(\ref{f14}). 
	
\end{itemize}


\section{Conclusion}\label{sec7}	
We now summarise our findings. First of all we obtain a static, charged black hole solution in the presence of perfect fluid dark matter (PFDM). Furthermore, we have incorporated the Newman-Janis algorithm in order to derive the metric corresponding to a rotating, charged black hole surrounded by PFDM. Our preliminary study on the event horizon radius of this black hole reveals that the PFDM parameter $\alpha$ creates a noticeable influence on both outer and inner event horizons of the black hole. We observe that there exists a certain value $\alpha_c$ (at a fixed value of spin $a$ and charge $Q$ of the black hole) up to which the outer event horizon radius ($r_{h+}$) decreases with the increase in the value of $\alpha$. However, after $\alpha_c$, surprisingly $r_{h+}$ starts to increase with increase in the value of $\alpha$. 
On the basis of this critical value, we define two ranges of values for $\alpha$. We speculate that this feature might be due to the presence of the PFDM contributing to the effective mass of the black hole system. Then we focus our study on the radius of photon spheres ($r_p$) for both prograde (co-rotating) and retrograde (counter-rotating) orbits. We observe that with the increase in $\alpha$, 
the radius $r_p$ decreases for $\alpha < \alpha_c$, and increases for $\alpha > \alpha_c$. Besides that we also observe that increase in the value of the spin parameter ($a$) increases the radius of the co-rotating photon orbits, however, the reverse is observed for the counter-rotating photon orbits. This has been observed for both $\alpha < \alpha_{c}$ and $\alpha > \alpha_{c}$ values of the PFDM parameter $\alpha$. On the other hand, increment in the value of the charge $Q$ of the black hole decreases the radius of the photon orbits for both co-rotating and counter-rotating case. Then we compute the energy ($E$) and angular momentum ($L$) for massive particles moving in prograde and retrograde orbits. The energy ($E$) of the particle in prograde orbits decreases whereas in retrograde orbits increases with increase in the radial distance ($r$) from the black hole and gets close to unity as the particle approaches infinity. The increment in the values of $\alpha$ and spin ($a$) increases the energy of the particle considerably for the prograde orbits. On the other hand in case of retrograde orbits, the energy ($E$) increases with the increase in $\alpha$ but falls with the increasing value of the spin ($a$) of the black hole. This is because when the particle spins along the black hole, the black hole helps its motion whereas in the reverse case it opposes. Furthermore, for a fixed value of spin $a$ and PFDM parameter $\alpha$, increase in the charge ($Q$) of the black hole results in decrease in the energy of the particle. A possible reason may be that for a fixed spin, the mass of the black hole increases with increase in the charge and hence the energy of the black hole increases which results in a decrease in the rotational energy of the particles. The most important observation is in the case of the magnitude of the angular momentum ($L$) of the particle which decreases with the increase in the value of $\alpha$ for $\alpha < \alpha_c$ and increases for $\alpha > \alpha_c$ for both types of orbits.

\noindent We have then studied charged particles. In case of charged particles we analyse both the energy ($E$) and angular momentum ($L$) with the variation in the charge $(q)$ of the particle and found that with the increase in the value of charge ($q$), the energy decreases in prograde orbits and increases in case of retrograde orbits. Besides that, we observe that angular momentum ($L$) increases with increasing $q$ for $\alpha < \alpha_c$ and falls for $\alpha > \alpha_c$. This feature is observed for both co-rotating and counter-rotating particles. It is observed that the effective potential of the black hole for photons and massive particles, increases with  increasing spin ($a$) and charge ($Q$) of the black hole, and also with the angular momentum ($L$) of the particle. 
Also with increase in the value of $\alpha$ ($\alpha < \alpha_c$), the potential increases whereas for $\alpha > \alpha_c$, it decreases. 

\noindent Finally, we studied the Penrose process. The negative energy particles are very important with respect to the idea of energy gain (extraction of energy) from black holes. We observed that negative energy considerably increases with increase in the negative angular momentum (counter-rotating particle). We find that the energy gain increases when there is an increment in the value of the PFDM parameter $\alpha$. The obtained expression for the energy gain also depends on the charge parameter $Q$. We find that increase in the charge parameter $Q$ decreases the energy gain for the outgoing particle.

\section*{Acknowledgments}

A.D. would like to acknowledge the financial support of S.N. Bose National Centre for Basic Sciences, Salt Lake City, Kolkata. A.S. acknowledges the financial support by Council of Scientific and Industrial Research (CSIR, Govt. of India). The authors would also like to thank the referees for very useful comments.

\appendix
\section*{Appendix A: Limiting case of the non-rotating charged black hole solution with perfect fluid dark matter}\label{AppA}
The black hole solution we have obtained in this paper has the following lapse function
\begin{eqnarray}
f(r)=1-\frac{2GM}{r}+\frac{Q^2}{r^2}+\frac{\alpha}{r}ln\Big(\frac{r}{|\alpha|}\Big).
\end{eqnarray}
 In the limit $\alpha \to 0$, we apply the L'Hopital rule to the term $-\frac{\ln(|\alpha|)}{\frac{r}{\alpha}}$ in the above equation. This leads to the Reissner-Nordstr\"{o}m black hole solution
\begin{eqnarray}
f(r)=1-\frac{2GM}{r}+\frac{Q^2}{r^2}~.
\end{eqnarray}
On the other hand, in the limit $\alpha \to 0$ and $Q \to 0$, we have
\begin{eqnarray}
f(r)=1-\frac{2GM}{r}
\end{eqnarray}
which corresponds to the Schwarzchild solution.
\section{Appendix B}
In this Appendix, we show the derivation of eq.(\ref{eq0}). We begin our calculation with the following expressions
\begin{eqnarray}
 F(r)=x^2 \Big(a^2 -\Delta\Big) + r^4 E^2 -2aEr^2 x -\Delta r^2=0~;~	E=\frac{1}{r^2 ax}\Big[\Big(a^2 -\Delta +\frac{r \Delta^{'}}{4}\Big)x^2 +\Big(\frac{\Delta^{'}}{4}r^3 -\frac{\Delta}{2}r^2\Big)\Big]~.
\end{eqnarray}
Hence we have $E^2$ as
\begin{eqnarray}
E^2=\frac{1}{r^4a^2x^2}\Bigg[\Big(a^2 -\Delta +\frac{r \Delta^{'}}{4}\Big)^2 x^4 + \Big(\frac{\Delta^{'}}{4}r^3 -\frac{\Delta}{2}r^2\Big)^2 + 2\Big(a^2 -\Delta +\frac{r \Delta^{'}}{4}\Big)\Big(\frac{\Delta^{'}}{4}r^3 -\frac{\Delta}{2}r^2\Big)x^2\Bigg].
\end{eqnarray}
Rearranging the expressions of $E$ and $E^2$, we obtain
\begin{eqnarray}\label{401}
r^2 a x E=\Big[\Big(a^2 -\Delta +\frac{r \Delta^{'}}{4}\Big)x^2 +\Big(\frac{\Delta^{'}}{4}r^3 -\frac{\Delta}{2}r^2\Big)\Big]
\end{eqnarray}
\begin{eqnarray}\label{402}
r^4a^2x^2E^2=\Bigg[\Big(a^2 -\Delta +\frac{r \Delta^{'}}{4}\Big)^2 x^4 + \Big(\frac{\Delta^{'}}{4}r^3 -\frac{\Delta}{2}r^2\Big)^2 + 2\Big(a^2 -\Delta +\frac{r \Delta^{'}}{4}\Big)\Big(\frac{\Delta^{'}}{4}r^3 -\frac{\Delta}{2}r^2\Big)x^2\Bigg].
\end{eqnarray}
Also multiplying $F(r)=0$ by $a^2x^2$, we obtain
\begin{eqnarray}\label{403}
a^2x^4 \Big(a^2 -\Delta\Big) + a^2x^2r^4 E^2 -2aEr^2 x a^2x^2 -\Delta a^2 x^2 r^2=0.
\end{eqnarray}
By substituting eq.(\ref{401}) and eq.(\ref{402}) in eq.(\ref{403}), we obtain the following expression
\begin{align}
a^2 x^4 \Big(a^2 -\Delta \Big) + \Big(a^2 -\Delta +\frac{r \Delta^{'}}{4}\Big)^2 x^4 + \Big(\frac{\Delta^{'}}{4}r^3 -\frac{\Delta}{2}r^2\Big)^2 + 2\Big(a^2 -\Delta +\frac{r \Delta^{'}}{4}\Big)\Big(\frac{\Delta^{'}}{4}r^3 -\frac{\Delta}{2}r^2\Big)x^2  \nonumber \\
-2a^2 x^2 \Bigg[ \Big(a^2 -\Delta +\frac{r \Delta^{'}}{4}\Big)x^2 +\Big(\frac{\Delta^{'}}{4}r^3 -\frac{\Delta}{2}r^2\Big)\Bigg] - \Delta a^2 x^2 r^2=0~.
\end{align}
The above expression has terms involving $x^4$, $x^2$ and $x^0$. So collecting the coefficients of these powers of $x$ leads to
\begin{align}
\Big[4\Big(\Delta -a^2- \frac{r \Delta^{'}}{4}\Big)^2 - 4a^2\Big(a^2 -\Delta  +\frac{r \Delta^{'}}{2}\Big)\Big]x^4 + \Big[\Big(4a^2 -4\Delta +r\Delta^{'}\Big) \nonumber \\
 	\times	\Big(\frac{r^3 \Delta^{'}}{2} -r^2 \Delta\Big)-2r^3 a^2 \Delta^{'}\Big]x^2 + \Big(r^2 \Delta -\frac{r^3 \Delta^{'}}{2}\Big)^2=0
 \end{align}
which is eq.(\ref{eq0}).

\end{document}